\documentclass[
reprint,
superscriptaddress,
longbibliography,
bibnotes,
amsmath,amssymb,
aps,
prb,
]{revtex4-2}
\usepackage[utf8]{inputenc}
\usepackage[english]{babel}
\usepackage[plainpages = false, pdfpagelabels, 
                 bookmarks,
                 bookmarksopen = true,
                 bookmarksnumbered = true,
                 breaklinks = true,
                 linktocpage,
                 colorlinks = true,
                 linkcolor = blue,
                 urlcolor  = blue,
                 citecolor = blue,
                 anchorcolor = green,
                 hyperindex = true,
                 hyperfigures
                 ]{hyperref} 
\usepackage{enumerate}
\usepackage{tabularx}
\usepackage{makecell, multirow}
\usepackage{tikz}
\usetikzlibrary{patterns}
\usepackage{slashed}
\usepackage{physics}
\usepackage{verbatim}
\usepackage{svg}
\usepackage{mathtools}
\usepackage{cancel}
\usepackage{bbold}
\usepackage{bm}
\usepackage{soul}
\usepackage{booktabs}
\usepackage[caption=false,singlelinecheck=off]{subfig}
\usepackage{blindtext}
\usepackage{bbm}
\usepackage{graphicx}
\usepackage{dcolumn}
\usepackage[normalem]{ulem}
\usepackage{xcolor}
\usepackage[percent]{overpic}

\renewcommand{\vec}[1]{\bm{#1}}

\begin{document}

\title{Topological signatures in the curvature-induced energy response}

\author{Jaehyeok Lee}
\affiliation{\mbox{Department of Physics, Konkuk University, Seoul 05029, Republic of Korea}}

\author{Iuegyun Hong}
\affiliation{\mbox{Department of Physics, Konkuk University, Seoul 05029, Republic of Korea}}

\author{Jinhong Park}
%\email{jinhongpark@konkuk.ac.kr}
\affiliation{\mbox{Department of Physics, Konkuk University, Seoul 05029, Republic of Korea}}

\date{\today}
\begin{abstract}
Relativistic effective field theory predicts a topological energy response to a gravitational field that appears at third order in spatial gradients. Here, we investigate how this response emerges in the nonrelativistic Haldane model using a microscopic lattice formulation of curvature-induced deformations. We find that the leading first-order energy response is nonuniversal and depends on the bond-resolved structure of the deformation; in particular, it vanishes for a symmetric modulation of the three nearest-neighbor hoppings. At third order, the curvature-induced energy response contains a bond-symmetric component together with nonuniversal bond-dependent contributions. The bond-symmetric component, which is isolated by a uniform modulation of the three hoppings, exhibits a universal discontinuity across the topological transition whose magnitude agrees with the relativistic gravitational Chern--Simons prediction. Thus, although the full curvature-induced energy response generally contains nonuniversal lattice contributions, its third-order response retains a universal topological component beyond the relativistic limit.
\end{abstract}
\maketitle
\section{Introduction}
Recent advances in thermal transport experiments have established heat currents as a powerful probe of strongly correlated topological quantum matter. In quantum Hall systems, measurements of the quantized thermal conductance have provided direct access to the net chirality of edge excitations and, through it, to information that is not contained in the electrical Hall conductance alone~\cite{KaneFisher1997,Cappelli2002}. Particularly striking examples are the observation of quantized heat flow carried by fractional quantum Hall edge modes~\cite{Banerjee2017} and the subsequent measurement of a half-integer thermal Hall conductance at filling factor $\nu = 5/2$~\cite{Banerjee2018}. These experiments have demonstrated that thermal transport can reveal fundamental characteristics of topological phases, including their chiral central charge and the nature of their edge degrees of freedom. Related ideas have also played an important role in candidate Kitaev spin liquids, where a half-quantized thermal Hall response has been discussed as a signature of chiral Majorana edge modes, with particular attention to the coupling between the topological edge mode and bulk phonons~\cite{Kasahara2018, VinklerAvivRosch2018, YeBalents2018, Klocke2022, RuCl3Ultraclean2025}. In particular, there have been extensive theoretical and experimental studies~\cite{Spanslatt2019, Dutta2022,DuttaInterface2022, Kumar2022,Srivastav2022,LeBreton2022, Hashisaka2023, Park2024, Paul2024}, which characterize topological order through edge transport. In comparison with these remarkable developments in transport, however, the static response of topological states to spatial deformations and gravitational fields remains much less explored, both experimentally and theoretically.

An important distinction between charge and energy responses already emerges at the level of {\it relativistic} effective field theory. For a two-dimensional Chern insulator, the leading transverse charge response follows from the electromagnetic Chern--Simons term and is first order in spatial gradients,
\begin{align}
J_{i}^C
 = -\frac{e^2 \mathcal{C}}{h}\epsilon_{ij}\partial_j\Phi\,, 
\end{align}
where $\mathcal{C}$ denotes the Chern number and $\Phi$ a scalar field. The corresponding gravitational response in a relativistic low-energy theory has a qualitatively different structure. The gravitational Chern--Simons action produces an energy-momentum response to a {\it gradient of curvature}~\cite{Stone2012}, given by
\begin{align}
J^E_{i} &= \frac{\hbar c^2 (c_R-c_L)}{96 \pi}\epsilon_{ij}\partial_j R \nonumber \\
&= - \frac{\hbar c^2 (c_R-c_L)}{48 \pi} \epsilon_{ij}\partial_j\nabla^2\psi , 
\end{align} 
where $\psi$ denotes a static gravitational potential, $c$ is the speed of light, $R = - 2 \nabla^2 \psi + O(\psi^2)$ is the scalar curvature and $c_R-c_L$ is the difference between right- and left-moving central charges characterizing the edge modes of the system. The leading topological energy response is therefore third order in spatial gradients, in sharp contrast to the first-order electrical Hall response.  

This unusual third-order gravitational response was demonstrated microscopically in the Haldane model in Ref.~\cite{Park2022}, where a spatially varying gravitational potential was found to generate a transverse energy current proportional to its third derivative. Importantly, that work also showed that such a gravitational response should not in general be identified with the response to a temperature gradient, emphasizing the distinction between static geometric response and thermal transport. Related recent work has further revisited the connection between inhomogeneous temperature profiles, curved spacetime, and gravitational anomalies~\cite{Bermond2024}.

Graphene subject to strain and smooth lattice deformations has been intensively studied as a platform for engineering its electronic properties, e.g., in Refs.~\cite{PhysRevLett.97.196804,PhysRevLett.97.016801,VOZMEDIANO2010109,jun2025nanowrinkle,kim2008graphene}. In particular, nonuniform strain can generate effective gauge fields and pseudomagnetic fields, providing a route to modify the low-energy Dirac spectrum without chemical doping. Out-of-plane deformation and local curvature have also been observed in graphene-based structures, as demonstrated experimentally in graphene nanobubbles~\cite{levy2010strain}. Moreover, such deformations can be locally controlled through the interaction between graphene and a scanning tunneling microscope tip~\cite{georgi2017tuning}. These developments motivate the study of electronic and energy responses induced directly by smooth curvature.

In this paper, we investigate how these distinct charge and energy responses emerge from a microscopic lattice deformation. We consider the Haldane model on a smoothly curved graphene sheet, where curvature modifies both the nearest-neighbor hoppings and an effective scalar potential. We find a striking contrast between the two responses. The leading transverse charge current is generated solely by the curvature-induced scalar potential and is fixed by the Chern number, while the hopping modulation does not contribute at this order. 

The energy response behaves qualitatively differently: the scalar potential gives no transverse energy current, whereas the curvature-induced hopping modulation generates a finite response. Its leading, first-gradient contribution is nonuniversal and depends on the bond-resolved microscopic structure of the deformation. At third order in spatial gradients, however, a remarkably different structure emerges. Although lattice and other nonrelativistic corrections render the absolute response nonuniversal away from the transition, the third-order coefficient exhibits a discontinuity across the topological transition whose magnitude agrees with the relativistic gravitational Chern--Simons prediction. Thus, a characteristic fingerprint of the relativistic gravitational response survives in the microscopic lattice system even beyond the regime in which the full response is described by a relativistic Dirac theory.

The remainder of the paper is organized as follows. In Sec.~\ref{sec:model}, we introduce the Haldane model on a smoothly curved graphene sheet and formulate the curvature-induced modifications of the nearest-neighbor hoppings and scalar potential within a microscopic Slater--Koster description. In Sec.~\ref{sec:chargeresponse}, we study the transverse charge-current response and show that its leading gradient contribution is determined by the Chern number and the spatial gradient of the curvature-induced scalar potential. In Sec.~\ref{sec:energyresponse}, we turn to the energy-current response. We first construct the microscopic energy-current operator from the lattice continuity equation and then evaluate its response systematically in a spatial-gradient expansion. We show that the leading linear-gradient contribution is nonuniversal and depends on the bond-resolved deformation, whereas the third-order response retains a characteristic discontinuity across the topological transition inherited from the relativistic gravitational response. We summarize our results and discuss possible experimental realizations in Sec.~\ref{sec:conclusion}.

We set $\hbar = e = 1$ throughout this paper. 

\section{Model: curved Haldane model}
\label{sec:model}

\subsection{Flat Haldane model}
\label{subsec:Flat model}
\begin{figure*}[t]
    \centering
    \includegraphics[width=0.92\textwidth]{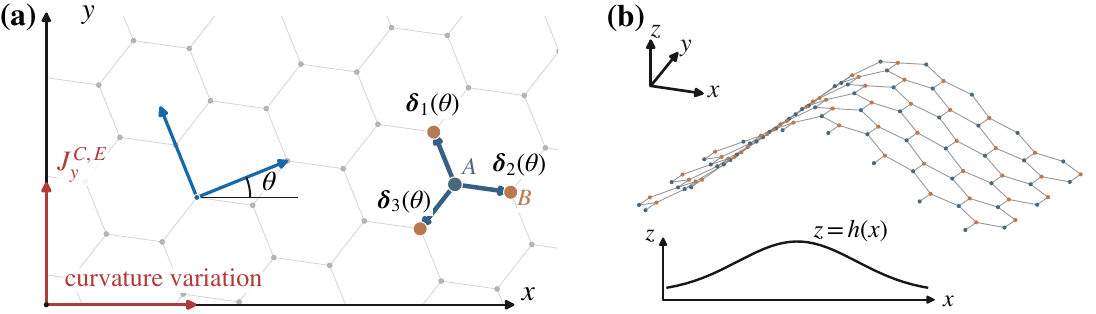}
    \caption{ (a) Flat graphene with crystal orientation specified by the angle $\theta$, defined as the tilt of the armchair direction relative to the laboratory $y$ axis. The deformation varies along the laboratory $x$ direction, while the transverse charge and energy currents, $J_y^C$ and $J_y^E$, flow along the $y$ direction. The three nearest-neighbor bond vectors $\vec{\delta}_\ell(\theta)$ ($\ell=1,2,3$) are indicated. (b) Three-dimensional view of graphene under smooth pure bending, described by the height profile $z=h(x)$ and translationally invariant along the $y$ direction. The inset shows the corresponding profile $z = h(x)$.}
    \label{fig:geometry_orientation}
\end{figure*}

We consider the Haldane model~\cite{Haldane1988} on a smoothly deformed graphene sheet. As a reference, we first introduce the flat system shown in Fig.~\ref{fig:geometry_orientation}(a), choosing the crystal orientation such that the armchair direction lies along the $y$ axis ($\theta=0$). The Hamiltonian consists of three terms:
\begin{align} \label{eq:flatHamiltonian}
    H_0 = H_{\text{NN}} + H_{\text{NNN}} + H_{\text{sp}}\,,
\end{align}
where
\begin{subequations}
\begin{align}
     H_{\text{NN}} &= \sum_{ (i, j) \in \text{NN}} \tau_{ij} a_{i}^\dagger b_{j} + \text{h.c.}, \\ 
     H_{\text{NNN}} &= i t_2 \sum_{ (i, j) \in \text{NNN}} (a_{i}^\dagger a_{j} - b_{i}^\dagger b_{j})+\text{h.c.} ,  \\ 
     H_{\text{sp}} & =  \sum_{i} m (a_{i}^\dagger a_{i} - b_{i}^\dagger b_{i})\,. 
\end{align}
\end{subequations}
Here $a_i$ ($a_i^\dagger$) and $b_i$($b_i^\dagger$) are the annihilation (creation) operators on the $A$ and $B$ sublattices, respectively. The term $H_{\text{NN}}$ describes hopping between nearest-neighbor (NN) sites with hopping amplitude $\tau_{ij}$. For flat graphene, the NN hopping is spatially uniform, $\tau_{ij} = -t_1 = V_{pp\pi}<0$, where $V_{pp\pi}$ denotes the $\pi$ bonding hopping amplitude between neighboring graphene $p_z$ orbitals. The term $H_{\text{NNN}}$ describes complex next-nearest-neighbor (NNN) hopping with amplitude $\pm i t_2$, where $t_2\in\mathbb{R}$. This term breaks time-reversal symmetry. The sum over NNN pairs is taken over one of the two possible orientations of each NNN bond, with the opposite orientation included through the Hermitian conjugate. Finally, $H_{\text{sp}}$ denotes the staggered potential, which generates opposite on-site energies $\pm m$ on the two sublattices. 

Owing to translational invariance, Eq.~\eqref{eq:flatHamiltonian} can be written in momentum space as
\begin{align} \label{eq:flatHaldanemodel}
    \mathcal{H}_0 (\vec{k}) = \vec{d}_0(\vec{k}) \cdot \vec{\sigma}\,,
\end{align}
where $\vec{\sigma}=(\sigma_x,\sigma_y,\sigma_z)^T$ denotes the Pauli-matrix vector and $\vec{d}_0(\vec{k})=(d_{0,x},d_{0,y},d_{0,z})^T$ is a real vector with components
\begin{subequations}
\begin{align}
    d_{0,x} (\vec{k}) &= - t_1 \sum_{\ell=1}^3 \cos (\vec{k}\cdot \vec{\delta}_{\ell})\,, \\ 
      d_{0,y} (\vec{k}) &= t_1 \sum_{\ell=1}^3 \sin (\vec{k}\cdot \vec{\delta}_{\ell}) \,, \\ 
      d_{0,z} (\vec{k}) & = m - 2 t_2 \sum_{\ell=1}^3 \sin (\vec{k}\cdot \vec{b}_{\ell})\,. 
\end{align}
\end{subequations}
The three NN bond vectors from an $A$-sublattice site to the neighboring $B$-sublattice sites [Fig.~\ref{fig:geometry_orientation}(a)] are $\vec{\delta}_1=a(0,1)^T$, $\vec{\delta}_2=a(-\tfrac{\sqrt{3}}{2},-\tfrac{1}{2})^T$, and $\vec{\delta}_3=a(\tfrac{\sqrt{3}}{2},-\tfrac{1}{2})^T$, where $a$ is the NN bond length. The NNN displacement vectors on a given sublattice are $\vec{b}_1=\vec{\delta}_2-\vec{\delta}_3$, $\vec{b}_2=\vec{\delta}_3-\vec{\delta}_1$, and $\vec{b}_3=\vec{\delta}_1-\vec{\delta}_2$.

A general crystal orientation is obtained by rotating the lattice through an angle $\theta$ [Fig.~\ref{fig:geometry_orientation}]. The bond vectors then transform as $\vec{\delta}_{\ell}(\theta)=R_{\theta}\vec{\delta}_{\ell}$ and $\vec{b}_{\ell}(\theta)=R_{\theta}\vec{b}_{\ell}$, where $R_{\theta}$ is the two-dimensional rotation matrix. Equivalently, one may keep the bond vectors fixed and make the replacement $\vec{k}\rightarrow \vec{k}_{\theta} \equiv R_{-\theta}\vec{k}$ in the Hamiltonian. Throughout, the $x$ and $y$ axes remain fixed in the laboratory frame.

\subsection{Curvature-modified Haldane model}
\label{subsec:curvature_hopping}

We next consider a smooth out-of-plane deformation described by the height profile $z=h(x)$, as illustrated in Fig.~\ref{fig:geometry_orientation}(b). The deformation varies only along the $x$ direction, so translational invariance along $y$ is preserved. We work in a pure-bending approximation in which changes in the NN carbon--carbon bond lengths are neglected. Curvature then induces slow spatial variations in the NN hopping amplitudes $\tau_{ij}$ and in the scalar potential.

\begin{figure}[h]
    \centering
    \includegraphics[width=0.5\textwidth]{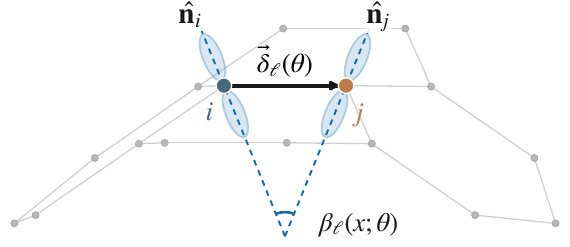}
    \caption{Slater--Koster model for a nearest-neighbor hopping in smoothly curved graphene. The highlighted sites $i\in A$ and $j\in B$ are connected by the $\ell$th bond vector $\vec{\delta}_\ell(\theta)$. The local $p_z$-orbital directions $\hat{\mathbf n}_i$ and $\hat{\mathbf n}_j$ are misaligned by the bond-dependent angle $\beta_\ell(x;\theta)$. This misalignment depends on the orientation of the bond relative to the bending direction and gives rise to the curvature-induced modulation of the nearest-neighbor hopping.}
    \label{fig:sk_geometry}
\end{figure}

We describe the curvature-induced NN hopping using the Slater--Koster parametrization~\cite{SlaterKoster1954}, illustrated in Fig.~\ref{fig:sk_geometry}. Denoting the local surface normals by $\hat{\mathbf n}_i$ and $\hat{\mathbf n}_j$, the hopping between NN sites~\cite{Isacsson2008} is
\begin{equation} 
    \tau_{ij} = V_{pp\pi}\, \hat{\mathbf n}_i\cdot\hat{\mathbf n}_j + (V_{pp\sigma}-V_{pp\pi}) (\hat{\mathbf n}_i\cdot\hat{\vec \delta}_{ij}) (\hat{\mathbf n}_j\cdot\hat{\vec \delta}_{ij}),
    \label{eq:sk_hopping}
\end{equation}
where $\vec{\delta}_{ij}=\vec r_j-\vec r_i$ is the three-dimensional NN bond vector and $\hat{\vec{\delta}}_{ij}\equiv\vec{\delta}_{ij}/|\vec{\delta}_{ij}|=(\vec r_j-\vec r_i)/a$ is its unit vector in the pure-bending approximation. The parameter $V_{pp\sigma}$ denotes the $\sigma$-bonding hopping amplitude between neighboring $p_z$ orbitals. For flat graphene, $\hat{\mathbf n}_i$ and $\hat{\mathbf n}_j$ are parallel and perpendicular to $\vec{\delta}_{ij}$, so the NN hopping reduces to $\tau_{ij}=V_{pp\pi}=-t_1<0$. In the presence of curvature, the relative orientation of neighboring $p_z$ orbitals produces spatially varying NN hopping amplitudes.

For a smooth deformation, the NN hopping amplitudes $\tau_{ij}$ vary slowly in space. The three bond-resolved NN hopping amplitudes can then be written as 
\begin{equation}
    \tau_\ell(x;\theta)
    =-t_1 + \delta\tau_\ell(x;\theta), \qquad \ell=1,2,3,
    \label{eq:bond_resolved_hopping}
\end{equation}
where \(\delta\tau_\ell(x;\theta)\) denotes the curvature-induced modulation of the hopping along the \(\ell\)th bond. We define the bond-orientation factor
\begin{equation} \label{eq:bondorientationfactor}
    p_\ell(\theta)= \frac{1}{a}|\hat{\mathbf y}\times \vec{\delta}_\ell(\theta)|\,. 
\end{equation} 
The hopping modulation is then
\begin{align} \label{eq:hoppingmodulation}
    \delta \tau_{\ell}(x;\theta)=\sin^2\!\Big(\frac{\beta_\ell(x;\theta)}{2}\Big) \left[ V_{pp\pi}\left(-2+p_\ell^2(\theta)\right) -V_{pp\sigma}p_\ell^2(\theta) \right].
\end{align}
Here, $\beta_\ell(x;\theta)$ is the relative angle between the local orbital directions across the $\ell$th bond. In the smooth-curvature limit, the normal-vector misalignment is determined by the projection of the bond onto the curvature direction and, to leading order, satisfies $\beta_\ell(x;\theta)\simeq a\,\kappa(x)\,p_\ell(\theta)$. For $\beta_\ell\ll1$, the hopping modulation reduces to
\begin{align}
    \delta \tau_{\ell}(x;\theta) \simeq \frac{a^2\kappa^2(x)p_\ell^2(\theta)}{4} \left[ V_{pp\pi}\left(-2+p_\ell^2(\theta)\right)-V_{pp\sigma}p_\ell^2(\theta) \right].
    \label{eq:weak_curvature}
\end{align}
Thus, all three hopping modulations share the same spatial dependence $\kappa(x) = h''(x)/[1+h'(x)^2]^{3/2}$, while their relative amplitudes are set by the bond-orientation factors $p_\ell(\theta)$.

\begin{figure}[t]
    \centering
    \includegraphics[width=\columnwidth]{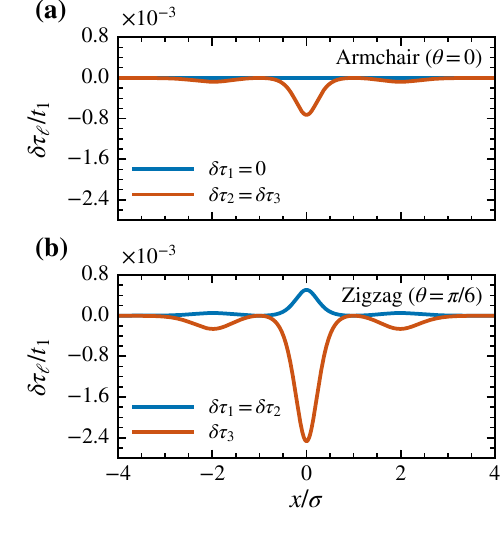}
    \caption {Curvature-induced nearest-neighbor hopping modulations $\delta\tau_\ell/t_1$ as functions of $x/\sigma$ for two representative crystal orientations: (a) armchair, $\theta=0$, and (b) zigzag, $\theta=\pi/6$. We use the Gaussian bending profile $h(x)=A e^{-x^2/(2\sigma^2)}$ with $A=3.7~\mathrm{nm}$ and $\sigma=2.5~\mathrm{nm}$, together with $t_1= -V_{pp\pi}=2.66~\mathrm{eV}$ and $V_{pp\sigma}=6.38~\mathrm{eV}$~\cite{Tomanek1988}. The three bond modulations can be obtained from Eq.~\eqref{eq:weak_curvature} by computing the curvature $\kappa(x)$ and the bond-orientation factors $p_\ell(\theta)$ in Eq.~\eqref{eq:bondorientationfactor}.}
    \label{fig:hopping_modulation}
\end{figure}

In addition to the NN hopping modulation, curvature induces an effective scalar potential through modifications of the NNN hoppings and $\pi$--$\sigma$ rehybridization between neighboring carbon orbitals~\cite{kim2008graphene}. On the lattice, this contribution is written as
\begin{align}
    H_{\text{scalar}}=\sum_i \Phi(\vec r_i)\big(a_i^\dagger a_i+b_i^\dagger b_i\big)\,.
    \label{eq:scalar_potential_lattice}
\end{align}
Although its magnitude can in principle be determined microscopically~\cite{kim2008graphene}, we treat the scalar potential phenomenologically.

Assuming that the curvature varies slowly in space, the curved lattice model can be described within a continuum approximation. Since translational invariance is preserved along the $y$ direction but broken along $x$, $k_y$ remains a good quantum number. The total Hamiltonian in this continuum approximation takes the form, in second quantization, as 
\begin{align} \label{eq:second_quantization}
  H_{\text{curv}} =  \frac{1}{L_x} \sum_{k_x,k_y,q_x}\int dx ~ e^{iq_x x}
    \vec{\Psi}_{\vec{k}_-}^\dagger \mathcal{H}(\vec{k},x; \theta) \vec{\Psi}_{\vec{k}_+}\,,
\end{align}
where $\vec{\Psi}_{\vec{k}} = (a_{\vec{k}}, b_{\vec{k}})^T$ is a two-component spinor in the sublattice space, and $q_x$ denotes the momentum transfer induced by the broken translational invariance along the $x$ direction. We parameterize the two momenta as 
$\vec{k}_{\pm} \equiv (k_x \pm q_x/2, k_y)$, respectively, where $\vec{k}$ is their average momentum. The local Hamiltonian $\mathcal{H}(\vec{k},x;\theta)$ then reads 
\begin{equation} 
    \mathcal{H}(\vec{k},x;\theta) = \mathcal{H}_0 (\vec{k}_{\theta}) + \delta \mathcal{H} (\vec{k},x;\theta) \,.
    \label{eq:local_hamiltonian}
\end{equation}
Here, the first term $\mathcal{H}_0$ is given in Eq.~\eqref{eq:flatHaldanemodel} for the flat Haldane model with the rotated crystal orientation by angle $\theta$, while the second term arises from the curvature-induced perturbation, 
\begin{subequations}
 \label{eq:curved_lattice_hamiltonian}
\begin{align}
   \delta \mathcal{H} (\vec{k},x;\theta) &= \delta \mathcal{H}_{\tau} (\vec{k},x;\theta) +  \delta \mathcal{H}_{\Phi} (\vec{k},x;\theta)\,, \\
\delta \mathcal{H}_{\tau} (\vec{k},x;\theta) &=\delta\vec d(\vec{k},x;\theta) \cdot\vec\sigma \,,
\label{eq:tauperturbation}\\ \delta \mathcal{H}_{\Phi} (\vec{k},x;\theta)
&=\Phi(x)\mathbbm{1} \,, \label{eq:phiperturbation}
\end{align}
\end{subequations}
where $\delta\vec d(\vec{k},x; \theta) = (\delta d_x,\delta d_y, 0)^T$ is given by 
\begin{subequations}
\begin{align}
    \delta d_x(\vec{k},x; \theta)&= \sum_{\ell=1}^{3} \delta\tau_\ell(x;\theta) \cos[\vec{k}\cdot\vec{\delta}_\ell(\theta)]\,,\\
    \delta d_y(\vec{k},x; \theta)&= -\sum_{\ell=1}^{3} \delta\tau_\ell(x;\theta)
    \sin[\vec{k}\cdot\vec{\delta}_\ell(\theta)]
    \label{eq:curvature_d_vector}\,. 
\end{align}
\end{subequations}
We use the convention in which the curvature-induced modulation $\delta \tau_{\ell}(x;\theta)$ [Eq.~\eqref{eq:hoppingmodulation}] of the NN hopping couples to the center of the corresponding bond. In this convention, the local Hamiltonian \eqref{eq:local_hamiltonian} has no explicit $q_x$ dependence. The momentum transfer $q_x$ enters through the momenta $\vec{k}_{\pm}$ of the spinor fields. The local Hamiltonian in Eq.~\eqref{eq:local_hamiltonian} induces the charge response and energy responses, discussed in Secs.~\ref{sec:chargeresponse} and \ref{sec:energyresponse}, respectively. 

The resulting hopping modulations $\delta\tau_\ell(x)$ are shown in Fig.~\ref{fig:hopping_modulation} for representative crystal orientations, illustrating that curvature modulates the three NN bonds with different amplitudes.

\section{Charge response}
\label{sec:chargeresponse}

We next consider the transverse charge current along the $y$ direction in response to the curvature along the $x$ direction, see Fig.~\ref{fig:geometry_orientation}. The local charge-current density operator at $x$ takes the form of 
\begin{align} \label{eq:chargecurrentop}
    \hat{J}^C_y (x) = \frac{1}{L_xL_y} \sum_{k_x,k_y, q_x} e^{iq_x x} 
    \vec{\Psi}_{\vec{k}_-}^\dagger \mathcal{J}_y^C (\vec{k},x;\theta) \vec{\Psi}_{\vec{k}_+}\,,    
\end{align}
with the charge-current vertex kernel,
\begin{align}
    \mathcal{J}_y^C (\vec{k}, x; \theta)  = \frac{\partial \mathcal{H}(\vec{k}, x; \theta)}{\partial k_y} =\mathcal{J}_{y,0}^C (\vec{k}_{\theta})+ \delta \mathcal{J}_y^C (\vec{k}, x; \theta)\,.
\end{align}
Substituting Eqs.~\eqref{eq:local_hamiltonian} and \eqref{eq:curved_lattice_hamiltonian}, the charge-current vertex kernel can be decomposed into the flat contribution and curvature-induced contribution as
\begin{subequations}
    \begin{align} \label{eq:flatcurrentvertex}
    &\mathcal{J}_{y,0}^C (\vec{k}_{\theta}) = 
    \frac{\partial \mathcal{H}_0(\vec{k}_{\theta})}{\partial k_y}, \\ &\delta \mathcal{J}_y^C (\vec{k}, x; \theta) =
    \frac{\partial (\delta \mathcal{H}(\vec{k}, x; \theta))}{\partial k_y} =  \frac{\partial (\delta \mathcal{H}_{\tau}(\vec{k}, x; \theta))}{\partial k_y} \,. \label{eq:curvaturecurrentvertex}
    \end{align}
\end{subequations}
Using the lesser Green function $G^<_{\vec{k}_+, \vec{k}_-} (\omega) \equiv i\langle  \vec{\Psi}_{\vec{k}_-}^\dagger (\omega) \vec{\Psi}_{\vec{k}_+} (\omega)\rangle $, the expectation value $J_y^C(x) = \langle \hat{J}^C_y (x) \rangle$ can be expressed as 
\begin{align} \label{eq:currentdensityexp}
    J_y^C(x) &= - i \frac{1}{L_xL_y} \sum_{k_x,k_y,q_x}  e^{iq_x x} \int \frac{d\omega}{2\pi} \nonumber \\ &\times
   \text{Tr}[\mathcal{J}_y^C (\vec{k},x;\theta) G^<_{\vec{k}_+, \vec{k}_-} (\omega)]\,. 
\end{align}
Here, the trace is performed over the sublattice space, and the average is taken over the eigenstates of the Hamiltonian, $\mathcal{H}(\vec{k},x;\theta)$ in Eq.~\eqref{eq:local_hamiltonian}. The lesser Green function $G^<$ can be obtained by expanding up to the first order in $\delta \mathcal{H}(\vec{k},x;\theta) $ as 
\begin{align} \label{eq:greenfunexpand}
    G_{\vec{k}_+, \vec{k}_-}^<(\omega) &\simeq g_{\vec{k}}^<(\omega) \delta_{q_x,0} + \frac{1}{L_x} \int dx_1 e^{-i q_x x_1}  \nonumber \\  \times & [g^R_{\vec{k}_+} (\omega) \delta\mathcal{H}(\vec{k},x_1;\theta) g_{\vec{k}_-}^<(\omega) 
    \nonumber \\ 
    &+g^<_{\vec{k}_+} (\omega) \delta\mathcal{H}(\vec{k},x_1;\theta)g_{\vec{k}_-}^A(\omega) ]\,. 
\end{align}
The $g$'s are the Green functions in the absence of curvature and have only one momentum argument due to translational invariance. They are given by 
\begin{subequations}
\begin{align} \label{eq:baregreenfun}
    g^{R/A}_{\vec{k}} (\omega) &= \frac{1}{\omega-\mathcal{H}_0 (\vec{k}_{\theta})\pm i \eta}\,, \\ g^{<}_{\vec{k}} (\omega) &= f_0 (\omega) [g^A_{\vec{k}}(\omega) -g^R_{\vec{k}}(\omega)] \,,
\end{align}
\end{subequations}
where $f_0 (\omega)$ is the Fermi function and $\eta \rightarrow 0^+$ is a positive infinitesimal. Plugging Eq.~\eqref{eq:greenfunexpand} into Eq.~\eqref{eq:currentdensityexp} and retaining terms up to first order in the curvature-induced perturbation, the charge current can be decomposed into
\begin{align}
    J_y^C(x) = J_{y,\text{eq}}^C + J_{y, \text{cont}}^C(x) + J_{y,\text{bub}}^C(x)\,.
\end{align}
Here $J_{y,\text{eq}}^C$ is the equilibrium current of the flat system
\begin{align}
    J_{y,\text{eq}}^C &= - \frac{2}{L_x L_y} \sum_{\vec{k}} \int \frac{d\omega}{2\pi} f_0 (\omega)\text{Im}\big[\text{Tr}[\mathcal{J}_{y,0}^C(\vec{k}_{\theta})g^{R}_{\vec{k}}(\omega)]\big]\,,
\end{align}
which vanishes upon integration over the first Brillouin zone. The second term $J_{y,\text{cont}}^{C}$ is the contact contribution in response to $\delta\mathcal J_y^C (\vec{k},x;\theta)$, and using Eq.~\eqref{eq:curvaturecurrentvertex}, it is given by
\begin{align} \label{eq:contacttermsimplied}
    J_{y,\text{cont}}^C(x) &= - \frac{2}{L_x L_y} \sum_{\vec{k}} \int \frac{d\omega}{2\pi} f_0 (\omega) 
    \nonumber \\ &
    \times \text{Im}\big[\text{Tr}[\frac{\partial \delta \mathcal{H}_{\tau}(\vec{k},x;\theta)}{\partial k_y} g^{R}_{\vec{k}}(\omega)]\big]\,.
\end{align}
The last term $J_{y,\text{bub}}^C(x)$ involves the bubble diagram, given by
\begin{align}
    J_{y,\text{bub}}^C(x) &= {\!-}\frac{2}{L_x^2 L_y} \sum_{k_x,k_y,q_x}\int \frac{d\omega}{2\pi}f_0 (\omega) \int dx_1 \text{Im}\big[e^{i q_x (x-x_1)} \nonumber \\ & \times \text{Tr}[\mathcal{J}_{y,0}^C(\vec{k}_{\theta})  g^R_{\vec{k}_+} \delta\mathcal{H} (\vec{k},x_1; \theta) g^R_{\vec{k}_-}] \big]\,.
\end{align}
This bubble contribution can be split into the two contributions by splitting $\delta\mathcal{H} (\vec{k},x_1; \theta)$ as in Eq.~\eqref{eq:curved_lattice_hamiltonian}: the one arising from the scalar potential $\Phi(x)$ and the other from the modulation of the NN hopping $\delta \tau (x)$, 
\begin{subequations}
\begin{align} \label{eq:bubblechargecurrent}
    J_{\text{bub}}^C (x) &= \delta J_{y, \Phi}^C(x)+ \delta J_{y, \tau}^C(x) \,,\\
    \delta J_{y, \Phi}^C(x) &= {\!-}\frac{2}{L_x^2 L_y} \sum_{k_x,k_y,q_x}\int \frac{d\omega}{2\pi}f_0 (\omega) \int dx_1 \nonumber \\ & \times \text{Im}\big[e^{i q_x (x-x_1)}\text{Tr}[ \mathcal{J}_{y,0}^C(\vec{k}_{\theta})  g^R_{\vec{k}_+} \delta\mathcal{H}_{\Phi} (\vec{k},x_1; \theta) g^R_{\vec{k}_-}] \big], \label{eq:bubblechargecurrentscalarpotential} \\ \delta J_{y, \tau}^C(x) &={\!-}\frac{2}{L_x^2 L_y} \sum_{k_x,k_y,q_x}\int \frac{d\omega}{2\pi} f_0 (\omega) \int dx_1 \nonumber \\ & \times \text{Im}\big[e^{i q_x (x-x_1)} \text{Tr}[\mathcal{J}_{y,0}^C(\vec{k}_{\theta}) g^R_{\vec{k}_+} \delta\mathcal{H}_{\tau} (\vec{k},x_1; \theta) g^R_{\vec{k}_-}] \big]\,. \label{eq:bubblechargecurrentmodulation}
\end{align}
\end{subequations}
To linear order in the curvature-induced perturbation, the bubble contributions involve only the flat charge-current in Eq.~\eqref{eq:flatcurrentvertex}; the curvature-induced correction to the current vertex is already accounted for by the contact term in Eq.~\eqref{eq:contacttermsimplied}. 

We evaluate the charge-current density $J_y^C(x)$ by combining the contribution in Eq.~\eqref{eq:contacttermsimplied} with the bubble contributions in Eq.~\eqref{eq:bubblechargecurrent}. For a smoothly varying curvature, the bubble diagrams are expanded in the momentum transfer $q_x$, and the frequency and momentum integrations are performed order by order. The details of the gradient expansion and the frequency integration are presented in Appendix~\ref{app:gradient_expansion}.

The detailed evaluation of the contact and bubble contributions is presented in Appendix~\ref{app:charge_response}. Here, we summarize the main results. At zeroth order in $q_x$, the contact contribution~\eqref{eq:contacttermsimplied} from the curvature-induced NN-hopping modulation exactly cancels the corresponding bubble contribution in Eq.~\eqref{eq:bubblechargecurrentmodulation}, while the scalar-potential bubble in Eq.~\eqref{eq:bubblechargecurrentscalarpotential} vanishes separately. Furthermore, the NN-hopping contribution to the bubble diagram vanishes to linear order in $q_x$. 

Consequently, to leading order in the spatial gradient, the transverse charge current originates solely from the curvature-induced scalar potential. Expanding the scalar-potential contribution to linear order in $q_x$ and performing the frequency integration, we obtain the zero-temperature charge-current density at position $x$, 
\begin{align} 
    J_y^C(x)
    &=
    -\partial_x\Phi(x)
    \int_{\mathrm{1BZ}}
    \frac{d^2k}{(2\pi)^2}\,
    \Omega_-(\vec{k})
    +\mathcal{O}(\partial_x^2)
    \nonumber\\
    &=
    -\frac{\mathcal{C}}{2\pi}\,
    \partial_x\Phi(x)
    +\mathcal{O}(\partial_x^2),
    \label{eq:charge_current_chern}
\end{align}
where $\Omega_-(\vec{k})$ is the Berry curvature of the occupied band,
\begin{align} \label{eq:Berrycurvature}
    \Omega_-(\vec{k})
    =
    i\,\Tr\left[
        P_-(\vec{k})
        \left[
            \partial_{k_x}P_-(\vec{k}),
            \partial_{k_y}P_-(\vec{k})
        \right]
    \right],
\end{align}
and $\mathcal{C}$ is the Chern number of the occupied band, 
\begin{align} \label{eq:chernnumberintegral}
    \mathcal{C}
    &=
    \frac{1}{2\pi}
    \int_{\mathrm{1BZ}} d^2k\,
    \Omega_-(\vec{k}) \nonumber \\
    &= \frac{1}{2} [\text{sgn}(m-3\sqrt{3}t_2) - \text{sgn}(m+3\sqrt{3}t_2)]\,.
\end{align}
Here $P_{\pm}(\vec{k}) \equiv |\vec{k},\pm\rangle \langle \vec{k}, \pm|$ denotes the projector onto the upper $+$ and lower $-$ bands at a given momentum $\vec{k}$, respectively.  Thus, the leading transverse charge response is determined entirely by the topological invariant of the flat Haldane model and the spatial gradient of the curvature-induced scalar potential. Accordingly, this leading contribution is independent of the crystal orientation angle $\theta$, reflecting its topological origin rather than the microscopic bond anisotropy.

\begin{figure}[h]
    \centering
    \includegraphics[width=0.9\columnwidth]{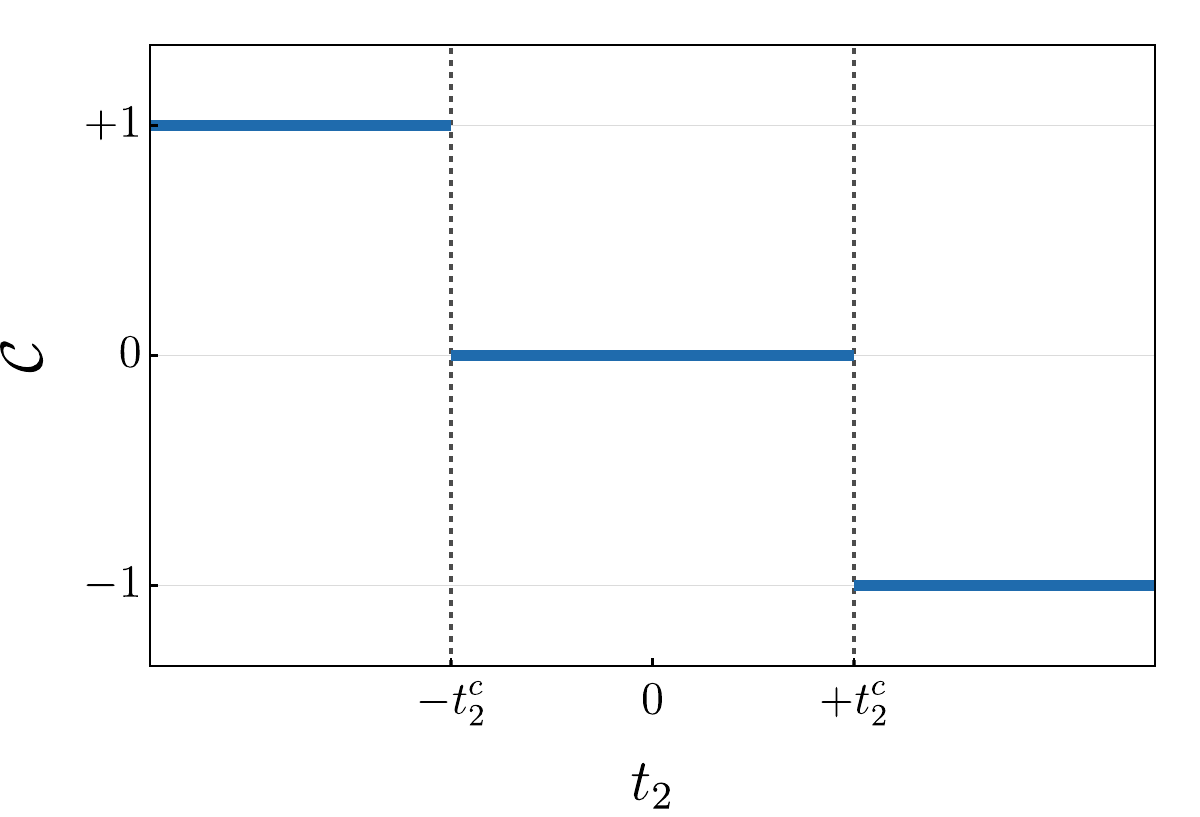}
    \caption{Chern number $\mathcal{C}$ of the occupied band as a function of the next-nearest-neighbor hopping $t_2$. It was obtained numerically by integrating the Berry curvature in Eq.~\eqref{eq:Berrycurvature} and performing the momentum integral in Eq.~\eqref{eq:chernnumberintegral}. The topological phase transitions occur at $t_2=\pm t_2^c$, where $t_2^c=m/(3\sqrt{3})$. We choose the parameters $t_1 = 2/3$ and $m = 5$ for computation.}
    \label{fig:chern_number}
\end{figure}

We compute the Berry curvature in Eq.~\eqref{eq:Berrycurvature} and perform the momentum integral in Eq.~\eqref{eq:chernnumberintegral} numerically. We plot the numerically obtained $\mathcal{C}$ as a function of $t_2$ with fixed $t_1 = 2/3$ and $m = 5$ in Fig.~\ref{fig:chern_number}. As a function of $t_2$, the system undergoes topological phase transitions at $t_2=\pm t_2^c$, with $t_2^c=m/(3\sqrt{3})$. The Chern number vanishes in the topologically trivial regime $|t_2|<t_2^c$ and takes the quantized values $\mathcal{C}=\pm1$ for $|t_2|>t_2^c$, with the sign determined by the sign of $t_2$. The curvature-induced scalar potential thus generates a transverse charge response governed directly by the Chern number. Consequently, the coefficient relating the transverse charge current to the gradient of the scalar potential changes discontinuously across the topological transition and is quantized within each gapped topological phase, providing a direct manifestation of the bulk topology in the curvature-induced response.

\section{Energy response}
\label{sec:energyresponse}
We now turn to the energy-current response. Unlike the charge current, the energy-current operator is defined from the continuity equation for the local energy density. To derive the energy-current operator in the presence of the spatially varying deformation, we introduce an auxiliary ``gravitational field"~\cite{Luttinger1964,Qin2011,Park2022} coupled to the local energy density. After deriving the corresponding current operator, the auxiliary field is set to zero. The details of the derivation are given in Appendix~\ref{app:energycurrent_derivation}.

\begin{figure*}[t] 
    \centering
    \includegraphics[width=0.9\textwidth]{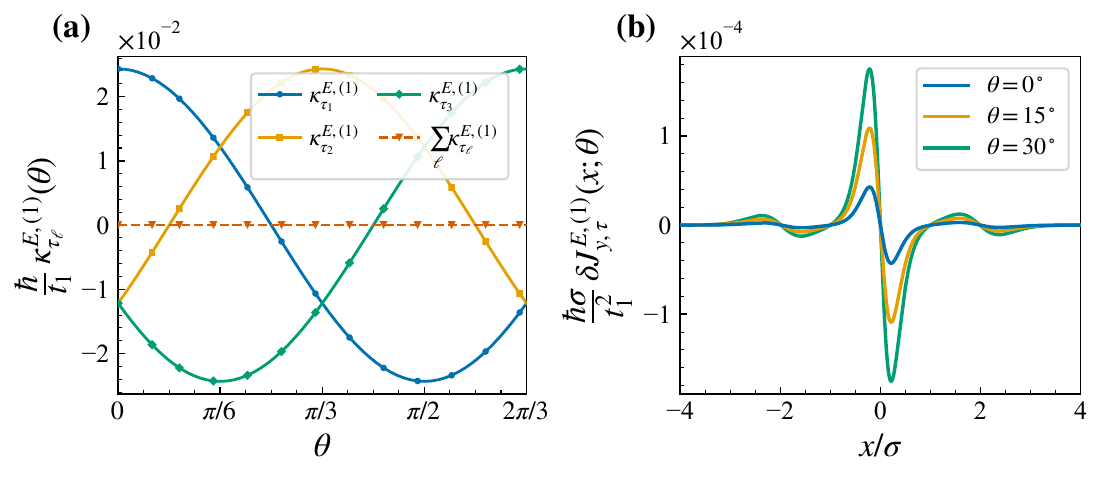}
    \caption{Transverse energy-current response generated by the first spatial derivative of the curvature-induced hopping modulation. (a) Bond-resolved response coefficients $\kappa_{\tau_\ell}^{E,(1)}(\theta)$ for $\ell=1,2,3$ as a function of the crystal orientation $\theta$. Although the individual contributions depend on $\theta$, their unweighted sum $\sum_{\ell}\kappa_{\tau_\ell}^{E,(1)}(\theta)$ (dashed curve) vanishes. (b) Real-space energy current $\delta J_{y,\tau}^{E,(1)}(x;\theta)$ generated by the first derivative of the curvature-induced hopping modulation for $\theta=0$, $\pi/12$, and $\pi/6$. The resulting current profile is antisymmetric, $\delta J_{y,\tau}^{E,(1)}(-x;\theta)=-\delta J_{y,\tau}^{E,(1)}(x;\theta)$. We use $t_1=1$, $t_2=1.5$, $m=5$, and an $800\times800$ Brillouin-zone grid. The curvature profile is $h(x)=A e^{-x^2/(2\sigma^2)}$ with $A=3.7\,{\rm nm}$ and $\sigma = 2.5 \,{\rm nm}$. $\kappa_{\tau_\ell}^{E,(1)}(\theta)$ and $\delta J_{y,\tau}^{E,(1)}(x;\theta)$ are plotted in units of $t_1/\hbar$ and $t_1^2 / (\hbar\sigma)$, respectively.}
    \label{fig:energy_response_linear}
\end{figure*}

The resulting energy-current density along the $y$ direction can be written as
\begin{align}
    \hat J_y^E(x)
    =
    \frac{1}{L_xL_y}
    \sum_{k_x,k_y,q_x}
    e^{iq_xx}
    \Psi^\dagger_{\vec k_-}
    \mathcal J_y^E(\vec k,q_x,x;\theta)
    \Psi_{\vec k_+},
    \label{eq:energy_current_operator}
\end{align}
where $\mathcal J_y^E(\vec k,q_x,x;\theta)$ denotes the single-particle energy-current vertex. In contrast to the charge-current vertex, the energy-current vertex generally depends explicitly on the transferred momentum $q_x$. We separate it into the contribution from the flat Hamiltonian and the correction induced by the deformation,
\begin{align}
    \mathcal J_y^E(\vec k,q_x,x;\theta)
    =
    \mathcal J_{y,0}^E(\vec k_\theta,q_x)
    +
    \delta\mathcal J_y^E(\vec k,q_x,x;\theta).
    \label{eq:energy_current_vertex_decomposition}
\end{align}

\begin{figure}[!t]
    \centering
    \includegraphics[width=0.95\columnwidth]{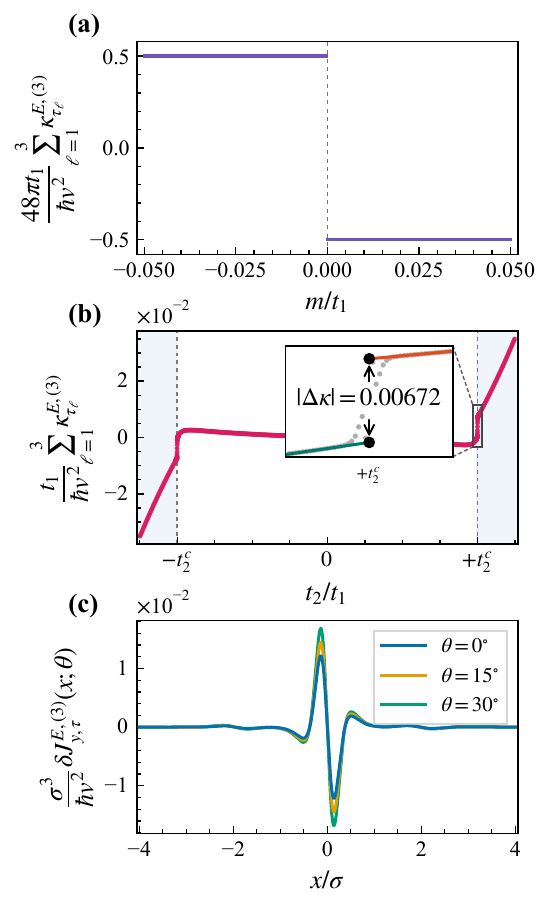}
    \caption{Transverse energy-current response generated by the third spatial derivative of the hopping modulation. (a) Bond-summed response coefficient $\sum_{\ell=1}^{3}\kappa_{\tau_\ell}^{E,(3)}$ in the single-Dirac-cone relativistic limit, plotted as a function of $m/t_1$. We set $t_2=0$, retain only the linear dispersion around the Dirac point $K$, and use $v=3 a t_1/2$. We plot $\sum_{\ell=1}^{3}\kappa_{\tau_\ell}^{E,(3)}$ in units of $\hbar v^2 /(48 \pi t_1)$, such that the response approaches $\pm 1/2$ as $m$ approaches zero from either side, and the resulting discontinuity at $m=0$ is unity.
    (b) Bond-response coefficient $\sum_{\ell=1}^{3}\kappa_{\tau_\ell}^{E,(3)}$ for an equal modulation of the three nearest-neighbor hoppings, $\delta\tau_1=\delta\tau_2=\delta\tau_3$, plotted as a function of $t_2/t_1$. The vertical dashed lines mark the two topological transition points $t_2=\pm t_2^c$, where $t_2^c=m/(3\sqrt{3})$. The shaded regions denote the topological phases $|t_2|>t_2^c$. The inset shows independent linear extrapolations of the two branches to the positive transition point, yielding $|\Delta\kappa^{E,(3)}|\simeq0.00672$ in units of ${\hbar}v^2/t_1$, close to the relativistic prediction $1/(48\pi)\simeq0.00663$.
    (c) Real-space third-order energy current $J_{y,\tau}^{E,(3)}(x;\theta)$ at $t_2/t_1=1.5$ for $\theta=0$, $\pi/12$, and $\pi/6$. The three crystal orientations exhibit the same spatial symmetry and line shape, with slightly different amplitudes. $J_{y,\tau}^{E,(3)}(x;\theta)$ is plotted in units of $\hbar v^2/\sigma^3$. Panels (b) and (c) use $m/t_1=5$, and an $800\times800$ Brillouin-zone grid. Panels (a) and (b) show $\theta=0$; both responses are independent of the crystal orientation $\theta$ (not shown).} 
    \label{fig:energy-response-q3}
\end{figure}

For the flat Hamiltonian $\mathcal H_0(\vec k_\theta)$ of Eq.~\eqref{eq:flatHaldanemodel}, the energy-current vertex in the long-wavelength regime is expanded up to third order as 
\begin{align}
    \mathcal J_{y,0}^E(\vec k_\theta,q_x)\simeq j_0^E(\vec k_\theta) + q_xj_1^E(\vec k_\theta) + q_x^2j_2^E(\vec k_\theta) + q_x^3j_3^E(\vec k_\theta).
    \label{eq:energy_current_flat_vertex}
\end{align}
The explicit expressions for the coefficients $j_n^E$ with $n=0, 1,2,3$ and their microscopic derivation are given in Appendix~\ref{app:energycurrent_derivation}.

The deformation also modifies the microscopic energy-current operator. The correction $\delta\mathcal J_y^E$ is obtained by applying the same construction to the deformed Hamiltonian $\mathcal H_0+\delta\mathcal H$ and retaining terms linear in the deformation. For the contact contribution discussed below, only its $q_x=0$ component is required. The corresponding expressions are given in Appendix~\ref{app:energycurrent_response}.

To linear order in the deformation, the expectation value of the energy current can be decomposed as
\begin{align}
    J_y^E(x;\theta) = J_{y,\mathrm{eq}}^E + J_{y,\mathrm{cont}}^E(x;\theta) + J_{y,\mathrm{bub}}^E(x;\theta).
    \label{eq:energycurrent_decomposition}
\end{align}
The equilibrium and contact contributions vanish after integration over the Brillouin zone for arbitrary crystal orientation $\theta$, as shown in Appendix~\ref{app:energycurrent_response}. Therefore, the energy-current response is determined solely by the bubble contribution,
\begin{align}
    J_y^E(x;\theta) = J_{y,\mathrm{bub}}^E(x;\theta), 
\end{align}
which is written as 
\begin{align}
    J_{y,\text{bub}}^E(x;\theta) &= {\!-}\frac{2}{L_x^2 L_y} \sum_{k_x,k_y,q_x}\int \frac{d\omega}{2\pi}f_0 (\omega) \int dx_1 \text{Im}\big[e^{i q_x (x-x_1)} \nonumber \\ 
    & \times \text{Tr}[ \mathcal{J}_{y,0}^E(\vec{k}_{\theta},q_x)  g^R_{\vec{k}_+} \delta\mathcal{H} (\vec{k},x_1; \theta) g^R_{\vec{k}_-}] \big]\,.
    \label{eq:bubbleenergycurrent}
\end{align}
As for the charge-current response, the bubble contribution can be split into the contributions arising from the scalar potential $\Phi(x)$ and the modulation of the NN hopping $\delta\tau(x)$,
\begin{subequations}
\begin{align}
    J_{\text{bub}}^E (x;\theta) &= \delta J_{y, \Phi}^E(x;\theta)+ \delta J_{y, \tau}^E(x;\theta) \,,\\
    \delta J_{y, \Phi}^E(x;\theta) &= {\!-}\frac{2}{L_x^2 L_y} \sum_{k_x,k_y,q_x}\int \frac{d\omega}{2\pi}f_0 (\omega) \int dx_1 \text{Im}\big[e^{i q_x (x-x_1)} \nonumber \\ 
    & \times \text{Tr}[ \mathcal{J}_{y,0}^E(\vec{k}_{\theta},q_x)  g^R_{\vec{k}_+} \delta\mathcal{H}_{\Phi} (\vec{k},x_1; \theta) g^R_{\vec{k}_-}] \big], 
    \label{eq:bubbleenergycurrentscalarpotential} \\
    \delta J_{y, \tau}^E(x;\theta) &={\!-}\frac{2}{L_x^2 L_y} \sum_{k_x,k_y,q_x}\int \frac{d\omega}{2\pi} f_0 (\omega) \int dx_1 \text{Im}\big[e^{i q_x (x-x_1)}  \nonumber \\ 
    & \times \text{Tr}[\mathcal{J}_{y,0}^E(\vec{k}_{\theta},q_x) g^R_{\vec{k}_+} \delta\mathcal{H}_{\tau} (\vec{k},x_1; \theta) g^R_{\vec{k}_-}] \big]\,.
    \label{eq:bubbleenergycurrentmodulation}
\end{align}
\end{subequations}

The bubble contribution is expanded systematically up to third order in $q_x$, as detailed in Appendix~\ref{app:energycurrent_response}. We define the energy-current response coefficients through the gradient expansion 
\begin{align}
    \delta J_{y,\Gamma}^E(x;\theta) = \sum_{n=0}^{3} \kappa_{\Gamma}^{E,(n)}(\theta) \partial_x^n\Gamma(x),
    \label{eq:energy_response_gradient}
\end{align}
with $\Gamma=\Phi,\tau_{\ell}$. The coefficients $\kappa_{\Gamma}^{E,(n)}(\theta)$ are obtained by expanding the bubble response in $q_x$ and performing the frequency and Brillouin-zone integrations, as described in Appendix~\ref{app:energycurrent_response}. For the deformations considered here, the even-order coefficients vanish; an analytical derivation of $\kappa_{\Gamma}^{E,(0)}(\theta)=0$ is given in Appendix~\ref{app:energycurrent_response}. We thus focus below on the linear and cubic responses, $n=1$ and $n=3$.

We first consider the leading nonvanishing contribution, $n=1$. Figure~\ref{fig:energy_response_linear}(a) shows the bond-resolved coefficients $\kappa_{\tau_\ell}^{E,(1)}(\theta)$ for the three NN bonds. Each contribution depends on the crystal orientation $\theta$, while their unweighted sum satisfies
\begin{align}
    \sum_{\ell=1}^{3} \kappa_{\tau_\ell}^{E,(1)}(\theta) =0
\end{align}
Thus, a spatially varying hopping that modulates all three bonds equally does not contribute to the response $ \delta J_{y,\tau}^{E,(1)}(x;\theta)$. 

The curvature-induced deformation considered here, however, does not modulate the three bonds equally. For a deformation varying along the $x$ direction, the hopping modulations $\delta\tau_\ell(x)$ acquire different spatial profiles for the three bond orientations $\ell$ as illustrated in Fig.~\ref{fig:hopping_modulation}. The linear response 
\begin{align}
    \delta J_{y,\tau}^{E,(1)}(x;\theta)
    =
    \sum_{\ell=1}^{3}
    \kappa_{\tau_\ell}^{E,(1)}(\theta)
    \partial_x\delta\tau_\ell(x;\theta),
    \label{eq:energy_linear_hopping}
\end{align}
therefore remains finite. The resulting real-space current profiles are shown in Fig.~\ref{fig:energy_response_linear}(b) for several crystal orientations. For the symmetric curvature profile considered here, $\delta J_{y,\tau}^{E,(1)}(x;\theta)$ is antisymmetric in $x$. The scalar-potential contribution vanishes, $\kappa_{\Phi}^{E,(1)}(\theta)=0$ and hence $\delta J_{y,\Phi}^{E,(1)}(x;\theta) = 0$. 

We next consider the third-order response, $n=3$. In the relativistic continuum limit, the transverse energy-current response induced by a slowly varying gravitational field is governed by the gravitational Chern--Simons response~\cite{Stone2012,Park2022}. For a single massive Dirac cone, this yields 
\begin{align}
    J_y^{E,(3)}
    =
    \frac{v^2}{96\pi}
    \operatorname{sgn}(m)\,
    \partial_x^3\psi,
    \label{eq:energy_cubic_relativistic}
\end{align}
where $v$ is the Dirac velocity and $\psi$ is the gravitational field. To compare this result with our calculation, we consider an equal modulation of the three NN hoppings,  $\delta\tau_1=\delta\tau_2=\delta\tau_3\equiv\delta\tau$, in which the lattice deformation reduces in the relativistic Dirac theory to the gravitational field  $\psi=-\delta\tau/t_1$. The third-order response in our lattice calculation is then given by 
\begin{align}
    \delta J_{y,\tau}^{E,(3)}(x;\theta)
    = \Big(\sum_{\ell=1}^3\kappa_{\tau_\ell}^{E,(3)} (\theta) \Big) \partial_x^3\delta\tau (x) ,
    \label{eq:energy_third_hopping}
\end{align}
where the Dirac response coefficient corresponds to 
\begin{align}
    \sum_{\ell=1}^3 \kappa_{\tau_\ell}^{E,(3)}(\theta) = - \frac{v^2}{96 \pi t_1} \text{sgn}(m)\,. 
\end{align}
Thus, the relativistic theory predicts a discontinuity
\begin{align}
    \left| \Delta \sum_{\ell=1}^{3} \kappa_{\tau_\ell}^{E,(3)}\right| = \frac{v^2}{48\pi t_1}
\end{align}
across $m = 0$. Figure~\ref{fig:energy-response-q3}(a) shows that our calculation for the linearized single-Dirac-cone Hamiltonian with $t_2 = 0$ reproduces this prediction. 

Owing to the rotational symmetry of the linearized Dirac spectrum, the response is independent of the crystal orientation $\theta$, as we have also verified numerically; we thus show only $\theta = 0$ in Fig.~\ref{fig:energy-response-q3}(a).

We next consider the full lattice model. As reported in Ref.~\cite{Park2022}, the nonrelativistic lattice corrections render the third-order response nonuniversal away from the transition. Nevertheless, as shown in Fig.~\ref{fig:energy-response-q3}(b), a clear discontinuity remains at the topological transition. A linear extrapolation of the two branches toward the transition gives $|\Delta \sum_{\ell=1}^3\kappa^{E,(3)}_{\tau_{\ell}}|\simeq 0.00672$, in remarkable agreement with the relativistic prediction $1/(48\pi)\simeq 0.00663$ in units of $v^2/t_1$. Thus, while the absolute value of the third-order response contains nonuniversal lattice contributions, its discontinuity retains a clear fingerprint of the relativistic Dirac response. 

For the curvature-induced deformation considered in this paper, the three NN hoppings are modulated differently. The resulting third-order real-space energy current,
\begin{align}
    \delta J_{y,\tau}^{E,(3)}(x;\theta)
    =
    \sum_{\ell=1}^{3}
    \kappa_{\tau_\ell}^{E,(3)}(\theta)
    \partial_x^3\delta\tau_\ell(x;\theta),
\end{align}
is shown in Fig.~\ref{fig:energy-response-q3}(c) for several crystal orientations. The resulting spatial profiles have the same qualitative form, with only a weak dependence of their amplitudes on the crystal orientation.

Unlike the linear-order response shown in Fig.~\ref{fig:energy_response_linear}(a), for which the bond-summed coefficient vanishes, the bond-summed third-order response, $\sum_{\ell=1}^3\kappa_{\tau_\ell}^{E,(3)}$, remains finite. Furthermore, it is independent of the crystal orientation $\theta$, as we have verified numerically. This makes the orientation dependence of the real-space third-order current relatively weak.

Finally, we find no transverse energy-current response induced by the scalar potential $\Phi$. In particular, $\kappa_{\Phi}^{E,(3)}(\theta)=0$, and more generally $\kappa_{\Phi}^{E,(n)}(\theta)=0$ for all orders considered in our gradient expansion. Thus, the curvature-induced transverse energy current discussed above originates entirely from the modulation of the nearest-neighbor hoppings.

\section{Conclusion and outlook}
\label{sec:conclusion}

In conclusion, we have studied the charge and energy responses induced by smooth curvature in the Haldane model using a microscopic lattice formulation. We adopt the Slater--Koster model to describe the curvature-induced modulation of the nearest-neighbor hoppings and the scalar potential. Within this framework, we systematically derive the transverse charge and energy responses. For the charge response, we recover the topological contribution associated with the Haldane transition, generated by the curvature-induced scalar potential.

For the energy response, we construct a local energy-current operator consistent with both the lattice continuity equation and the locality condition. Using this microscopic current operator, we show that the leading linear-gradient response is generally nonuniversal and depends on the bond-resolved structure of the deformation, whereas the third-order response exhibits a discontinuity across the topological transition whose magnitude agrees with the relativistic single-Dirac-cone prediction. Thus, although the absolute lattice energy response contains nonuniversal corrections, its discontinuity retains a clear fingerprint of the underlying relativistic topology. We further find that the scalar potential does not contribute to the transverse energy-current response at any order in the gradient expansion examined here. The curvature-induced transverse energy current therefore originates entirely from the modulation of the nearest-neighbor hoppings. Our results thus establish a direct connection between microscopic lattice deformations and topological charge and energy responses beyond the strict relativistic limit.

Compared with the charge response, the curvature-induced energy-current response is generally more challenging to access experimentally. Nevertheless, ultracold atoms in optical lattices provide a promising platform for probing such responses. The Haldane model has already been realized using ultracold fermions in a periodically driven honeycomb lattice~\cite{Jotzu2014}, while recent quantum-gas-microscope experiments have demonstrated local measurements of current and kinetic-energy operators with single-bond resolution~\cite{Impertro2024}. These developments suggest that the spatially modulated hopping deformation and the microscopic energy-current response considered here may be accessible in future programmable quantum simulators.

\section*{Acknowledgements}
J.~L., I.~H., and J.~P. acknowledge support from Korea NRF (Grant No. RS-2026-25492880).

\appendix

\section{Gradient expansion of Green functions}
\label{app:gradient_expansion}

We begin by considering the gradient expansion used to evaluate the charge- and energy-current responses. Since the deformation varies only along the $x$ direction, the momentum transfer takes the form $\vec q=(q_x,0)$. As defined in the main text, the two momenta connected by the momentum transfer $\vec{q}$ are $\vec{k}_{\pm}=\left(k_x\pm\frac{q_x}{2},k_y\right)$. The retarded Green functions of the flat system are
\begin{align}
    g_{\pm}^{R}(\omega)
    \equiv
    g_{\vec{k}_{\pm}}^{R}(\omega)
    =
    \frac{1}{
    \omega-\mathcal{H}_0(\vec{k}_{\pm,\theta})+i\eta
    }\,, 
\end{align}
with $\vec{k}_{\pm,\theta} \equiv R_{-\theta} \vec{k}_{\pm}$. In this and subsequent appendices, we use the shorthand notation $g_{\pm}^R \equiv g_{\pm}^{R}(\omega)$, suppressing the argument $\omega$ for simplicity. Similarly, we use the shorthand notation $g^R_{\vec{k}} (\omega) \equiv g^R$ for the bare Green function defined in Eq.~\eqref{eq:baregreenfun}. 

For a slowly varying deformation, the Green functions can be expanded in powers of $q_x$ as
\begin{align}
    g_{\pm}^{R} = g^{R} \pm q_x g_1^{R} +q_x^2 g_2^{R} \pm q_x^3 g_3^{R} +\mathcal{O}(q_x^4)\,. 
    \label{eq:green_gradient_expansion}
\end{align}
To express the expansion coefficients in a compact form, we introduce
\begin{align} \label{eq:Hderivatives}
    \mathcal H &\equiv
    \mathcal{H}_0(\vec{k}_{\theta}),
    \quad
    \mathcal H_x \equiv
    \frac{\partial \mathcal{H}_0(\vec{k}_{\theta})}{\partial k_x}, \nonumber \\ 
    \quad
    \mathcal H_{xx} &\equiv
    \frac{\partial^2 \mathcal{H}_0(\vec{k}_{\theta})}{\partial k_x^2},
    \quad
    \mathcal H_{xxx} \equiv
    \frac{\partial^3 \mathcal{H}_0(\vec{k}_{\theta})}{\partial k_x^3}.
\end{align}
Other momentum derivatives are denoted analogously, e.g., $\mathcal H_{xy}\equiv \partial_{k_x}\partial_{k_y}\mathcal H_0(\vec k_\theta)$ and $\mathcal H_{xxy}\equiv \partial_{k_x}^2\partial_{k_y}\mathcal H_0(\vec k_\theta)$. Taking the derivative $(g^R)^{-1} g^R = \mathbb{1}$ in $k_x$ in both sides, one can find a useful identity:
\begin{align} \label{eq:identity}
    \partial_{k_x} g^{R} = g^{R}\mathcal{H}_xg^{R},
\end{align}
\onecolumngrid
Using this identity \eqref{eq:identity}, we obtain
\begin{subequations}
\label{eq:green_gradient_coefficients}
\begin{align}
    g_1^{R}
    &=
    \frac{1}{2}
    g^{R}\mathcal H_x g^{R},
    \\
    g_2^{R}
    &=
    \frac{1}{8}
    g^{R}\mathcal H_{xx}g^{R}
    +
    \frac{1}{4}
    g^{R}\mathcal H_x g^{R}\mathcal H_x g^{R},
    \\
    g_3^{R}
    &=
    \frac{1}{48}
    g^{R}\mathcal H_{xxx}g^{R}
    +
    \frac{1}{16}
    g^{R}\mathcal H_{xx}g^{R}\mathcal H_x g^{R}
    +
    \frac{1}{16}
    g^{R}\mathcal H_x g^{R}\mathcal H_{xx}g^{R}
    +
    \frac{1}{8}
    g^{R}\mathcal H_x g^{R}\mathcal H_x g^{R}\mathcal H_x g^{R}.
\end{align}
\end{subequations}

We next consider a generic perturbation vertex $\Gamma(\vec{k})$ that has no explicit $q_x$ dependence. This applies to the scalar-potential and NN-hopping perturbations in the bond-center convention used in the main text. The product entering the bubble diagrams can then be expanded as
\begin{align}
    g_+^{R}\Gamma g_-^{R} = B_0 + q_x B_1 + q_x^2 B_2 + q_x^3 B_3 +\mathcal{O}(q_x^4).
    \label{eq:B_gradient_expansion}
\end{align}
The first four coefficients are
\begin{subequations}
\label{eq:B_gradient_coefficients}
\begin{align}
    B_0
    &=
    g^{R}\Gamma g^{R},
    \\
    B_1
    &=
    \frac{1}{2}
    \left(
    g^{R}\mathcal H_x g^{R}\Gamma g^{R}
    -
    g^{R}\Gamma g^{R}\mathcal H_x g^{R}
    \right),
    \\
    B_2
    &=
    \frac{1}{8}
    \left(
    g^{R}\mathcal H_{xx}g^{R}\Gamma g^{R}
    +
    g^{R}\Gamma g^{R}\mathcal H_{xx}g^{R}
    \right)
    +
    \frac{1}{4}
    \left(
    g^{R}\mathcal H_x g^{R}\mathcal H_x g^{R}\Gamma g^{R}
    +
    g^{R}\Gamma g^{R}\mathcal H_x g^{R}\mathcal H_x g^{R}
    -
    g^{R}\mathcal H_x g^{R}\Gamma g^{R}\mathcal H_x g^{R}
    \right),
    \\
    B_3
    &=
    \frac{1}{48}
    \left(
    g^{R}\mathcal H_{xxx}g^{R}\Gamma g^{R}
    -
    g^{R}\Gamma g^{R}\mathcal H_{xxx}g^{R}
    \right)
    \nonumber\\
    &\quad+
    \frac{1}{16}
    \Big(
    g^{R}\mathcal H_{xx}g^{R}\mathcal H_x g^{R}\Gamma g^{R}
    +
    g^{R}\mathcal H_x g^{R}\mathcal H_{xx}g^{R}\Gamma g^{R}
    -
    g^{R}\mathcal H_{xx}g^{R}\Gamma g^{R}\mathcal H_x g^{R}
    \nonumber\\
    &\qquad \qquad \qquad 
    +
    g^{R}\mathcal H_x g^{R}\Gamma g^{R}\mathcal H_{xx}g^{R}
    -
    g^{R}\Gamma g^{R}\mathcal H_{xx}g^{R}\mathcal H_x g^{R}
    -
    g^{R}\Gamma g^{R}\mathcal H_x g^{R}\mathcal H_{xx}g^{R}
    \Big)
    \nonumber\\
    &\quad+
    \frac{1}{8}
    \Big(
    g^{R}\mathcal H_x g^{R}\mathcal H_x g^{R}\mathcal H_x g^{R}\Gamma g^{R}
    -
    g^{R}\mathcal H_x g^{R}\mathcal H_x g^{R}\Gamma g^{R}\mathcal H_x g^{R}
    +
    g^{R}\mathcal H_x g^{R}\Gamma g^{R}\mathcal H_x g^{R}\mathcal H_x g^{R}
    -
    g^{R}\Gamma g^{R}\mathcal H_x g^{R}\mathcal H_x g^{R}\mathcal H_x g^{R}
    \Big).
\end{align}
\end{subequations}
\twocolumngrid
Equations~\eqref{eq:B_gradient_expansion}--\eqref{eq:B_gradient_coefficients} provide a systematic expansion of the response kernel in the momentum transfer $q_x$ as shown in the following subsections. 

\section{Frequency integration in the two-band model}
\label{appsubsec:frequencyintegration}

In this appendix, we perform the frequency integrals appearing in these expressions by using the spectral projectors of the two-band Hamiltonian. 

We begin by introducing a compact notation for the frequency integrals that appear in the gradient expansion of the response functions: 
\begin{align}
    \omega_N(A_1,\ldots,A_N)
    & \equiv 
    -2\int\frac{d\omega}{2\pi}\,
    f_0(\omega)\, \nonumber \\ & \times 
    \operatorname{Im}
    \Tr\left[
        A_1 g^R A_2 g^R \cdots A_N g^R
    \right],
    \label{eq:omegaN_definition}
\end{align}
where $A_1,\ldots,A_N$ are frequency-independent matrices acting in the sublattice space. The retarded Green function of the flat two-band system can be written in the spectral representation as
\begin{align}
    g^R
    =
    \frac{P_-}{\omega-E_-(\vec{k})+i\eta}
    +
    \frac{P_+}{\omega-E_+(\vec{k}) +i\eta},
    \label{eq:green_spectral}
\end{align}
where $E_\pm (\vec{k})$ are energies of the lower and upper bands at a given momentum $\vec{k}$, respectively. We further introduce the band gap at momentum $\vec{k}$, 
\begin{align}
    \Delta \equiv E_+(\vec{k})-E_-(\vec{k})>0.
\end{align}

Substituting Eq.~\eqref{eq:green_spectral} into Eq.~\eqref{eq:omegaN_definition}, the frequency integration can be performed analytically. We assume zero temperature and set the electrochemical potential in the middle of the gap, $\mu = 0$. The result can be expressed as
\begin{align}
    &\omega_N(A_1,\ldots,A_N)
    =  \sum_{r=1}^{N-1} 
      \sum_{\substack{s_1,\ldots,s_N=\pm\\
    N_-(s_1,\ldots,s_N)=r}}
   \frac{
        (-1)^{N+r-2}
    }{
        \Delta^{N-1}
    } \begin{pmatrix}
        N - 2\\ r-1
    \end{pmatrix} \nonumber \\ & \qquad \qquad \qquad
    \times \Tr\left[
        A_1 P_{s_1}
        A_2 P_{s_2}
        \cdots
        A_N P_{s_N}
    \right],
    \label{eq:omegaN_projector}
\end{align}
where $r = N_-(s_1,\ldots,s_N)$ denotes the number of occupied-band projectors $P_-$ in the sequence $\{P_{s_1},\ldots,P_{s_N}\}$.

The terms containing only occupied-band projectors or only unoccupied-band projectors vanish. The corresponding frequency integrals involve derivatives of the Fermi function $f_0(\omega)$ evaluated at the band energies, which vanish at zero temperature when the electrochemical potential lies inside the gap. At finite temperature, these contributions are  exponentially suppressed provided that $\Delta \gg k_B T$. Thus, only terms containing both occupied- and unoccupied-band projectors contribute to $\omega_N$, and consequently to the response.

For example, the frequency integral containing two Green functions reduces to
\begin{align}
    \omega_2(A_1,A_2)
    =
    -\frac{1}{\Delta}
    \Big\{
        \Tr\left[A_1 P_- A_2 P_+\right]
        +
        \Tr\left[A_1 P_+ A_2 P_-\right]
    \Big\}.
    \label{eq:omega2}
\end{align}
Similarly, for three Green functions, we obtain
\onecolumngrid
\begin{align}
    \omega_3(A_1,A_2,A_3)
   =
    \frac{1}{\Delta^2}
    &\Big\{
        \Tr\left[A_1 P_- A_2 P_+ A_3 P_+\right]
        +
        \Tr\left[A_1 P_+ A_2 P_- A_3 P_+\right]
 +
        \Tr\left[A_1 P_+ A_2 P_+ A_3 P_-\right]
    \nonumber \\    
    & -
        \Tr\left[A_1 P_+ A_2 P_- A_3 P_-\right]
        -
        \Tr\left[A_1 P_- A_2 P_+ A_3 P_-\right]
        -
        \Tr\left[A_1 P_- A_2 P_- A_3 P_+\right]
    \Big\}.
    \label{eq:omega3}
\end{align}
\twocolumngrid

Equation~\eqref{eq:omegaN_projector} allows us to evaluate the frequency integrals at arbitrary order in the gradient expansion without performing the frequency integration separately for each term. In particular, Eqs.~\eqref{eq:omega2} and \eqref{eq:omega3} are sufficient for the charge-current response discussed below, while higher-order functions $\omega_N$ will be used for the energy-current response.

\section{Charge-current response}
\label{app:charge_response}

We now apply the gradient expansion and the frequency-integration formulas in Appendices \ref{app:gradient_expansion} and \ref{appsubsec:frequencyintegration} to the charge-current response. 

We first consider the leading contribution, $q_x =0$, induced by the curvature,  starting with the NN-hopping perturbation $\delta\mathcal{H}_\tau(\vec{k},x;\theta)$ in Eq.~\eqref{eq:tauperturbation}. As discussed in Sec.~\ref{sec:chargeresponse}, there are two contributions to the charge response: the contact contribution \eqref{eq:contacttermsimplied} and the bubble contribution \eqref{eq:bubblechargecurrent}. The contact contribution \eqref{eq:contacttermsimplied} is written as 
\begin{align}
    J_{y,\mathrm{cont}}^C(x) = \int_{\mathrm{1BZ}} \frac{d^2k}{(2\pi)^2}\,
    \Tr\left[
        P_-\partial_{k_y} \delta \mathcal{H}_\tau (\vec{k},x;\theta)
    \right].
    \label{eq:charge_contact_appendix}
\end{align}
On the other hand, the zeroth-order bubble contribution from the same perturbation is
\begin{align}
    \delta J_{y,\tau}^{C,(0)}(x)
    =
    \int_{\mathrm{1BZ}}
    \frac{d^2k}{(2\pi)^2}\,
    \omega_2
    \left(
        \mathcal{J}_{y,0}^C,
        \Gamma_\tau
    \right).
\end{align}
Using Eq.~\eqref{eq:omega2} and the identity
\begin{align}
    \langle \vec{k}_{\theta}, \mp | \frac{\partial \mathcal{H}_0 (\vec{k}_\theta)}{\partial k_y} |\vec{k}_{\theta},\pm  \rangle = \pm \Delta  \langle \vec{k}_{\theta}, \mp |\frac{\partial }{\partial k_y} |\vec{k}_{\theta}, \pm\rangle\,, 
\end{align}
one finds
\begin{align}
   & \omega_2
    (
        \mathcal{J}_{y,0}^C,
        \delta \mathcal{H}_\tau (\vec{k}, x; \theta) 
   )
    =
    \Tr\left[
        (\partial_{k_y}P_-)\delta \mathcal{H}_\tau (\vec{k}, x; \theta) 
    \right].
    \label{eq:omega2_projector_derivative}
\end{align}
Consequently, the contact and zeroth-order bubble contributions combine into a total derivative in momentum space,
\begin{align}
    &J_{y,\mathrm{cont}}^C(x)
    +
    \delta J_{y,\tau}^{C,(0)}(x)
    =
    \int_{\mathrm{1BZ}}
    \frac{d^2k}{(2\pi)^2}\,
    \partial_{k_y}
    \Tr\left[
        P_-\delta \mathcal{H}_\tau (\vec{k}, x; \theta) 
    \right]
    \nonumber\\
    &\qquad \qquad \qquad \qquad \quad \,\,\, =0.
    \label{eq:charge_q0_tau_cancel}
\end{align}
This cancellation follows from the periodicity of the Brillouin zone.

For the scalar-potential perturbation in Eq.~\eqref{eq:phiperturbation}, there is no contact contribution since the scalar potential has no momentum dependence. Its zeroth-order bubble contribution also vanishes,
\begin{align}
    \delta J_{y,\Phi}^{C,(0)}(x)
    &=
    \Phi(x)
    \int_{\mathrm{1BZ}}
    \frac{d^2k}{(2\pi)^2}\,
    \omega_2
    \left(
        \mathcal{J}_{y,0}^C,\mathbbm{1}
    \right)
    =0,
    \label{eq:charge_q0_phi}
\end{align}
because $P_-P_+=P_+P_-=0$.

We next expand the bubble contributions to linear order in $q_x$. Using Eq.~\eqref{eq:B_gradient_expansion}, the linear-order response kernel for a generic perturbation $\Gamma$ is
\begin{align}
    \mathcal{K}^{C,(1)}_{\Gamma}(\vec{k})
    =
    \frac{1}{2}
    \left[
        \omega_3
        \left(
            \mathcal{J}_{y,0}^C,
            \mathcal{H}_x,
            \Gamma
        \right)
        -
        \omega_3
        \left(
            \mathcal{J}_{y,0}^C,
            \Gamma,
            \mathcal{H}_x
        \right)
    \right],
    \label{eq:charge_linear_kernel}
\end{align}
with $\mathcal{H}_x$ given in Eqs.~\eqref{eq:Hderivatives}.

For the NN-hopping modulation, all three matrices $\mathcal{J}_{y,0}^C$, $h_x$, and $\delta \mathcal{H}_\tau (\vec{k}, x; \theta) $ are traceless matrices in the two-band sublattice space. Using Eq.~\eqref{eq:omega3} together with the Pauli-matrix algebra, one finds
\begin{align}
    \omega_3
    \left(
        \mathcal{J}_{y,0}^C,
        \mathcal{H}_x,
        \delta \mathcal{H}_\tau (\vec{k}, x; \theta)
    \right)
    =
    \omega_3
    \left(
        \mathcal{J}_{y,0}^C,
        \delta \mathcal{H}_\tau (\vec{k}, x; \theta),
       \mathcal{H}_x
    \right).
\end{align}
Therefore, the NN-hopping contribution vanishes identically at linear order,
\begin{align}
    \delta J_{y,\tau}^{C,(1)}(x)=0.
    \label{eq:charge_linear_tau_zero}
\end{align}

The scalar-potential contribution, in contrast, remains finite. Setting $\Gamma=\mathbbm{1}$ in Eq.~\eqref{eq:charge_linear_kernel}, we obtain
\begin{align}
    \mathcal{K}^{C,(1)}_{\Phi}(\vec{k})
    &=
    \frac{1}{2}
    \left[
        \omega_3
        \left(
            \mathcal{J}_{y,0}^C,
            \mathcal{H}_x,
            \mathbbm{1}
        \right)
        -
        \omega_3
        \left(
            \mathcal{J}_{y,0}^C,
            \mathbbm{1},
            \mathcal{H}_x
        \right)
    \right]
    \nonumber\\
    &=
    -i\,\Omega_-(\vec{k}),
    \label{eq:charge_linear_phi_kernel}
\end{align}
where $\Omega_-(\vec{k})$ is the Berry curvature of the occupied band
\begin{align}
    \Omega_-(\vec{k})
    =
    i\,\Tr\left[
        P_-
        \left[
            \partial_{k_x}P_-,
            \partial_{k_y}P_-
        \right]
    \right]\,. 
\end{align}
Finally, using the identity, 
\begin{align}
    \frac{1}{L_x}
    \sum_{q_x}
    \int dx_1\,
    q_x e^{iq_x(x-x_1)}
    \Phi(x_1)
    =
    -i\,\partial_x\Phi(x),
\end{align}
the linear-gradient scalar-potential contribution becomes
\begin{align}
    \delta J_{y,\Phi}^{C,(1)}(x)
    &=
    -\partial_x\Phi(x)
    \int_{\mathrm{1BZ}}
    \frac{d^2k}{(2\pi)^2}\,
    \Omega_-(\vec{k})
    \nonumber\\
    &=
    -\frac{\mathcal{C}}{2\pi}
    \partial_x\Phi(x),
    \label{eq:charge_linear_phi_final}
\end{align}
with the Chern number 
\begin{align}
    \mathcal{C}
    =
    \frac{1}{2\pi}
    \int_{\mathrm{1BZ}}d^2k\,
    \Omega_-(\vec{k})
\end{align}
Combining
Eqs.~\eqref{eq:charge_q0_tau_cancel}, \eqref{eq:charge_q0_phi}, \eqref{eq:charge_linear_tau_zero}, and \eqref{eq:charge_linear_phi_final}, we recover the charge-current response,  Eq.~\eqref{eq:charge_current_chern} given in the main text.

\section{Derivation of the energy-current operator}
\label{app:energycurrent_derivation}

In this appendix, we derive the energy-current operator following the lattice construction of Ref.~\cite{Park2022}. To keep track of the physical positions of the lattice orbitals, we label an orbital by a composite index $i=(\mathbf R_i,\alpha_i)$, where $\alpha_i=A,B$, and denote its physical position by $\mathbf r_i=\mathbf R_i+\boldsymbol{\tau}_{\alpha_i}$ with $\boldsymbol{\tau}_{A} = (0,0)^T$ and $\boldsymbol{\tau}_{B} = \vec{\delta}_1 = a(0,1)^T$. The flat Haldane Hamiltonian can be decomposed into local energy operators 
\begin{align}
      \hat H_0
    =
    \sum_i \hat h_i,
    \qquad
    \hat h_i
    \equiv
    \frac{1}{2}\sum_j \hat X_{ij},
    \label{eq:local_energy_flat}
\end{align}
with the symmetrized operators $\hat{X}_{ij}$
\begin{align}
    \hat X_{ij}
    =
    c_i^\dagger \mathcal H^0_{ij} c_j
    +
    c_j^\dagger \mathcal H^0_{ji} c_i ,
    \label{eq:bondenergy}
\end{align}
where $\mathcal H^0_{ij}$ is the real-space matrix element of the flat Hamiltonian, and $\hat X_{ij}=\hat X_{ji}$ denotes the energy associated with the bond connecting orbitals $i$ and $j$.

Following Refs.~\cite{Luttinger1964,Qin2011,Park2022}, we introduce an auxiliary gravitational field $\psi$. Here we adopt a microscopic prescription of Ref.~\cite{Park2022}, where the gravitational field is coupled to each microscopic bond at its physical center, $\overline{\vec{r}}_{ij} \equiv \frac{\vec{r}_i+\vec{r}_j}{2}$. The corresponding local-energy operator is then given by
\begin{align}
    \hat h_i^\psi
    =
    \frac{1}{2}\sum_j
    u(\overline{\vec{r}}_{ij})\,\hat X_{ij},
    \qquad
    u(\vec{r})\equiv 1+\psi(\vec{r}).
    \label{eq:localenergy}
\end{align}
The gravitationally perturbed Hamiltonian is therefore
\begin{align}
    \hat H^\psi
    =
    \sum_i\hat h_i^\psi
    =
    \frac{1}{2}\sum_{i,j}
    u(\overline{\vec{r}}_{ij})\,\hat X_{ij}.
    \label{eq:Hpsi_real}
\end{align}
The discrete energy current from orbital $i$ to orbital $j$ is then defined through the local continuity equation,
\begin{align}
    \frac{d\hat h_i^\psi}{dt}
    +
    \sum_j \hat J_{ij}^{E,\psi}
    =0,
\end{align}
which gives
\begin{align}
    \hat J_{ij}^{E,\psi}
    &=
   i
    \big[
        \hat h_i^\psi,
        \hat h_j^\psi
    \big]
    =
   \sum_{lm}u(\overline{\vec{r}}_{il})
  u(\overline{\vec{r}}_{jm})
    \hat J_{ij;lm}^{E},
    \label{eq:bondenergycurrent_psi}
\end{align}
where
\begin{align}
    \hat J_{ij;lm}^{E}
    =
    \frac{i}{4}
    \big[
        \hat X_{il},
        \hat X_{jm}
    \big].
    \label{eq:bondenergycurrent}
\end{align}

To construct a continuous energy-current density from the discrete microscopic currents, we introduce a smooth function $f(\vec r)$ localized around $\vec r=0$ and normalized as
\begin{align}
    \int d^2r\, f(\vec r)=1.
\end{align}
The smoothing function varies on a length scale $\ell_f$ that is much longer than the lattice constant $a$, but much shorter than the length scale $\ell_\psi$ over which the gravitational field varies, $a\ll \ell_f\ll \ell_\psi$. The coarse-grained local energy density is defined by
\begin{align} 
    \hat h^{\psi}(\vec r)
    =
    \sum_i
    f(\vec r-\vec r_i)\,
    \hat h_i^{\psi}.
    \label{eq:smoothenergydensity}
\end{align}
Using the discrete continuity equation and the antisymmetry $\hat J_{ij}^{E,\psi}=-\hat J_{ji}^{E,\psi}$, the time derivative of Eq.~\eqref{eq:smoothenergydensity} can be written as
\begin{align}
    \frac{d\hat h^\psi(\vec r)}{dt}
    =
    -\frac{1}{2}\sum_{i,j}
    \left[
        f(\vec r-\vec r_i)
        -
        f(\vec r-\vec r_j)
    \right]
    \hat J_{ij}^{E,\psi}.
    \label{eq:continuity_smooth}
\end{align}
To express Eq.~\eqref{eq:continuity_smooth} as a spatial divergence, we use the exact identity
\begin{align}
    f(\vec r-\vec r_i)-f(\vec r-\vec r_j)
    &=
    \vec\nabla\cdot
    \left[
        \vec\delta_{ij}\,
        F_{ij}(\vec r)
    \right],
    \label{eq:f_identity}
\end{align}
where
\begin{align}
    F_{ij}(\vec r)
    &\equiv
    \int_{-1/2}^{1/2}d\lambda\,
    f\!\left(
        \vec r-\bar{\vec r}_{ij}
        -\lambda\vec\delta_{ij}
    \right),
    \qquad
    \vec\delta_{ij}\equiv\vec r_j-\vec r_i .
    \label{eq:Fij}
\end{align}
It follows that the coarse-grained energy-current density satisfying the continuity equation $\partial_t\hat h^\psi+\vec\nabla\cdot\hat{\vec J}^{E}_{\psi}=0$ is
\begin{align}
    \hat{\vec{J}}^{E}_{\psi}(\vec r)
    =
    \frac{1}{2}
    \sum_{i,j,l,m}
    \vec{\delta}_{ij}
    F_{ij}(\vec r)\,
    u(\bar{\vec r}_{il})
    u(\bar{\vec r}_{jm})
    \hat J_{ij;lm}^{E}.
    \label{eq:Jpsi_exact}
\end{align}
The factor $\vec{\delta}_{ij} = \vec r_j-\vec r_i$ in Eq.~\eqref{eq:Jpsi_exact} converts the energy-transfer rate between two orbitals into a spatial energy current.

We next exploit the separation of length scales $\ell_f\ll\ell_\psi$ to expand the slowly varying gravitational field. For this purpose, we define a point on the line connecting orbitals
$i$ and $j$ as
\begin{align}
    \vec r_{ij}(\lambda)
    =
    \bar{\vec r}_{ij}
    +
    \lambda\vec\delta_{ij},
    \qquad
    -\frac12\leq\lambda\leq\frac12 .
    \label{eq:rijlambda}
\end{align}
Using Eq.~\eqref{eq:Fij}, Eq.~\eqref{eq:Jpsi_exact} can then be written as
\begin{align}
    \hat{\vec J}^{E}_{\psi}(\vec r)
    &=
    \frac{1}{2}
    \sum_{i,j,l,m}
    \vec\delta_{ij}\,
    \hat J_{ij;lm}^{E}
    \nonumber \\ 
    &\times \int_{-1/2}^{1/2}d\lambda\,
    f\!\left(\vec r-\vec r_{ij}(\lambda)\right) u(\bar{\vec r}_{il})
    u(\bar{\vec r}_{jm}).
    \label{eq:Jpsi_line}
\end{align}
The two gravitational fields in Eq.~\eqref{eq:Jpsi_line} are evaluated at bond centers that in general differ from the point $\vec r_{ij}(\lambda)$ at which the current density is distributed. It is useful to introduce
\begin{align}
    \vec c_{ij;lm}
    \equiv
    \frac{
        \vec r_l+\vec r_m-\vec r_i-\vec r_j
    }{4},
    \label{eq:cijlm}
\end{align}
for which
\begin{align}
    \frac12
    \left[
        \bar{\vec r}_{il}
        +
        \bar{\vec r}_{jm}
    \right]
    -
    \vec r_{ij}(\lambda)
    =
    \vec c_{ij;lm}
    -
    \lambda\vec\delta_{ij}.
    \label{eq:center_difference}
\end{align}
We expand the product of the two slowly varying gravitational fields about the point $\vec r_{ij}(\lambda)$. To first order in spatial gradients, we obtain
\begin{align}
    u(\bar{\vec r}_{il})
    u(\bar{\vec r}_{jm})
    \approx
    &u^2\!\left(\vec r_{ij}(\lambda)\right)
    \nonumber\\
    &+
    \left(
        \vec c_{ij;lm}
        -
        \lambda\vec\delta_{ij}
    \right)
    \cdot
    \vec\nabla
    u^2\!\left(\vec r_{ij}(\lambda)\right).
    \label{eq:u_gradient_expansion}
\end{align}

Since $f(\vec r-\vec r_{ij}(\lambda))$ restricts $\vec r_{ij}(\lambda)$ to a region of size $\ell_f$ around $\vec r$, while $u(\vec r)$ varies on the much longer scale $\ell_\psi$, we may replace $u^2(\vec r_{ij}(\lambda))$ and $\nabla u^2(\vec r_{ij}(\lambda))$ by their values at $\vec r$. Using Eq.~\eqref{eq:u_gradient_expansion}, the energy current in Eq.~\eqref{eq:Jpsi_line} can be expanded to first order in gradients of the gravitational fields as 
\begin{align}
    \hat J_{\psi,a}^E(\vec r)
    =
    u^2(\vec r)\,
    \hat J_{a,\mathrm{loc}}^E(\vec r)
    +
    (\partial_b u^2(\vec r))\,
    \hat C_{ab}(\vec r),
    \label{eq:Jpsi_raw_expansion}
\end{align}
where
\begin{subequations}
\begin{align}
    \hat J_{a,\mathrm{loc}}^E(\vec r)
    &=
    \frac{1}{2}
    \sum_{i,j,l,m}
    \delta_{ij,a}\,
    \hat J_{ij;lm}^E
    \int_{-1/2}^{1/2}d\lambda\,
    f\!\left(
        \vec r-\vec r_{ij}(\lambda)
    \right),
    \label{eq:Jraw_real}
    \\
    \hat C_{ab}(\vec r)
    &=
    \frac{1}{2}
    \sum_{i,j,l,m}
    \delta_{ij,a}\,
    \hat J_{ij;lm}^E
    \int_{-1/2}^{1/2}d\lambda\,
    \nonumber \\ 
    & \quad \times \left(
        c_{ij;lm,b}
        -
        \lambda\delta_{ij,b}
    \right)
    f\!\left(
        \vec r-\vec r_{ij}(\lambda)
    \right),
    \label{eq:Cab_real}
\end{align}
\end{subequations}
with $a, b \in \{x,y\}$. Here, $\hat{J}^E_{a,\text{loc}}$ denotes the local contribution obtained by treating the gravitational fields as spatially uniform on the scale $\ell_{f}$, where $\hat{C}_{ab}$ accounts for the leading correction due to their spatial variation. The continuity equation determines the energy current only up to a divergence-free contribution. We note that Eq.~\eqref{eq:Jpsi_raw_expansion} is general and no symmetry of the tensor $\hat C_{ab}$ is assumed.

We now specialize to the transverse response considered in this work, where the perturbation occurs along the $x$ direction and the energy current is measured along $y$. In this case, Eq.~\eqref{eq:Jpsi_raw_expansion} reduces to
\begin{align}
    \hat J_{\psi,y}^E(\vec r)
    =
    u^2(\vec r)\hat J_{y,\mathrm{loc}}^E(\vec r)
    +
    (\partial_xu^2(\vec r))\hat C_{yx}(\vec r).
\end{align}
We fix this freedom by imposing the locality condition of Refs.~\cite{Qin2011,Park2022}: the energy-current operator in the presence of a slowly varying gravitational field must locally scale as $u^2(\vec r)$. The energy current $\hat{J}_{\psi,y}^E (\vec{r})$ is purely local in $u^2 (\vec{r})$ by adding the divergence-free contribution $-\partial_x [u^2(\vec{r}) \hat{C}_{yx}(\vec{r})]$ for the one-dimensional deformation case. This prescription leads to the expression 
\begin{align}
    \hat J_{\psi,y}^{E}(\vec r)
    =
    u^2(\vec r)\hat J_y^E(\vec r),
\end{align}
with 
\begin{align} \label{eq:energycurrentoperatorappendix}
    \hat J_y^E(\vec r)
    =
    \hat J_{y,\mathrm{loc}}^E(\vec r)
    -
    \partial_x\hat C_{yx}(\vec r).
\end{align}
Equation~\eqref{eq:energycurrentoperatorappendix} determines the flat-space transverse energy-current density for a slowly varying one-dimensional perturbation and can be applied directly to any lattice model.

We now transform the energy-current density operator, $\hat{J}_y^E(\vec{r})$, in Eq.~\eqref{eq:energycurrentoperatorappendix} to momentum space. We begin with the operator $\hat{J}_{y,\text{loc}}^E(\vec{r})$ in Eq.~\eqref{eq:Jraw_real}. The operator $\sum_{lm}\hat{J}^E_{ij;lm}$ entering Eq.~\eqref{eq:Jraw_real} can be written as the local energy operators $\hat{h}_i$ in Eq.~\eqref{eq:local_energy_flat} as 
\begin{align}
    \sum_{l,m}\hat J^E_{ij;lm}
    &=
    \frac{i}{4}
    \sum_{l,m}
    [\hat X_{il},\hat X_{jm}]
    =
    i[\hat h_i,\hat h_j].
    \label{eq:Jij_hihj}
\end{align}
Inserting the expression into Eq.~\eqref{eq:Jraw_real} and performing the Fourier transformation, we find that 
\begin{align}
    \hat J_{y,\mathrm{loc}}^E(\vec q)
    = 
    \frac{i f_{\vec{q}}}{2}
    \int_{-1/2}^{1/2}d\lambda
    \sum_{i,j}
    \delta_{ij,y}\,
    e^{-i\vec q\cdot\vec r_{ij}(\lambda)}
    [\hat h_i,\hat h_j].
    \label{eq:Jloc_hihj}
\end{align}
For a fixed $\lambda$, we introduce
\begin{align}
    \vec Q_1
    =
    \Big(\frac{1}{2}-\lambda\Big)\vec q,
    \qquad
    \vec Q_2
    =
    \Big(\frac{1}{2}+\lambda\Big)\vec q,
    \label{eq:Q1Q2}
\end{align}
so that
$
    e^{-i\vec q\cdot\vec r_{ij}(\lambda)}
    =
    e^{-i\vec Q_1\cdot\vec r_i}
    e^{-i\vec Q_2\cdot\vec r_j}.
$
The additional displacement $\delta_{ij,y}=r_{j,y}-r_{i,y}$ can be generated by introducing an auxiliary momentum $\vec s$,
\begin{align}
    \left.
    \partial_{s_y}
    \left[
    e^{-i(\vec Q_1+\vec s)\cdot\vec r_i}
    e^{-i(\vec Q_2-\vec s)\cdot\vec r_j}
    \right]
    \right|_{\vec s=0}
    =
    i\delta_{ij,y}
    e^{-i\vec q\cdot\vec r_{ij}(\lambda)}.
    \label{eq:auxiliary_momentum}
\end{align}
Defining the Fourier component of the local energy as
\begin{align}
    \hat h_{\vec q}
    \equiv
    \sum_i e^{-i\vec q\cdot\vec r_i}\hat h_i,
\end{align}
Eq.~\eqref{eq:Jloc_hihj} then becomes
\begin{align}
    \hat J_{y,\mathrm{loc}}^E(\vec q)
    =
    \frac{f_{\vec{q}}}{2}
    \int_{-1/2}^{1/2}d\lambda\,
    \left.
    \partial_{s_y}
    \left[
        \hat h_{\vec Q_1+\vec s},
        \hat h_{\vec Q_2-\vec s}
    \right]
    \right|_{\vec s=0}.
    \label{eq:Jloc_hqcommutator}
\end{align}
The Fourier component $\hat h_{\vec q}$ can be expressed as
\begin{align}
    \hat h_{\vec q}
    =
    \sum_{\vec k}
    \Psi_{\vec k_-}^\dagger
    D(\vec k_-,\vec k_+)
    \Psi_{\vec k_+},
    \label{eq:hq_momentum}
\end{align}
where $\vec k_\pm
    =
    \vec k\pm\frac{\vec q}{2}$, and the energy kernel $D(\vec k_1,\vec k_2)$
\begin{align}
    D(\vec k_1,\vec k_2)
    =
    \frac{1}{2}
    \left[
        \mathcal H_0(\vec k_1)
        +
        \mathcal H_0(\vec k_2)
    \right].
    \label{eq:Dkernel}
\end{align}
Using the canonical fermionic anticommutation relations, the commutator of two arbitrary Fourier components of the local energy is
\begin{align}
    [\hat h_{\vec Q},\hat h_{\vec Q'}]
    =
    \sum_{\vec k}
    \Psi_{\vec k_-}^\dagger
    F(\vec k;\vec Q,\vec Q')
    \Psi_{\vec k_+},
    \label{eq:hq_hq_commutator}
\end{align}
with 
\begin{align}
F(\vec{k}; \vec{Q}, \vec{Q}') =
    &
    D\big(
             \vec k_-,
        \vec k+\frac{\vec Q-\vec Q'}{2}
    \big)
    D\big(
      \vec k+\frac{\vec Q-\vec Q'}{2},
        \vec k_+
    \big)
    \nonumber\\
    -
    &
    D\big(
       \vec k_-,
        \vec k+\frac{\vec Q'-\vec Q}{2}
    \big)
    D\big(
         \vec k+\frac{\vec Q'-\vec Q}{2},
        \vec k_+
    \big)\,. 
\end{align}
Substituting $\vec Q\rightarrow\vec Q_1+\vec s$ and $\vec Q'\rightarrow\vec Q_2-\vec s$ into Eq.~\eqref{eq:hq_hq_commutator}, taking the derivative with respect to $s_y$, and changing $\lambda\rightarrow-\lambda$ in one of the two terms, we obtain
\begin{align}
    \mathcal J_{y,\mathrm{loc}}^E(\vec k,\vec q)
    = 
    \int_{-1/2}^{1/2}d\lambda\,
    \left.
    \partial_{p_y}
    \left[
        D(\vec k_-,\vec p)
        D(\vec p,\vec k_+)
    \right]
    \right|_{\vec p=\vec k+\lambda\vec q}.
    \label{eq:Jloc_momentum}
\end{align}
Here we use the fact that the coarse-graining function $f_{\vec{q}}\approx 1$ for a sufficiently slowly varying deformation. 

We next evaluate the correction $\hat C_{yx}(\vec r)$ in Eq.~\eqref{eq:Cab_real} following the same Fourier-transform procedure, Eqs.~\eqref{eq:Jij_hihj}–\eqref{eq:Jloc_momentum}. It is useful to separate the two contributions originating from $c_{ij;lm,x}$ and $-\lambda\delta_{ij,x}$ as
\begin{align}
    \mathcal C_{yx}(\vec k,\vec q)
    =
    \mathcal C_{yx}^{(c)}(\vec k,\vec q)
    +
    \mathcal C_{yx}^{(\lambda)}(\vec k,\vec q).
    \label{eq:Cyx_split}
\end{align}
For the first contribution $\mathcal C_{yx}^{(c)}(\vec k,\vec q)$, we use
\begin{align}
    c_{ij;lm,x}
    =
    \frac{1}{4}
    \left(
        \delta_{il,x}+\delta_{jm,x}
    \right)
\end{align}
and introduce the first spatial moment of the local energy,
\begin{align}
    \hat P_{i,x}
    =
    \frac{1}{4}
    \sum_l
    \delta_{il,x}\hat X_{il}.
    \label{eq:Pix}
\end{align}
The corresponding microscopic current contribution can then be expressed as
\begin{align}
    \sum_{l,m}
    c_{ij;lm,x}\hat J^E_{ij;lm}
    =
    \frac{i}{2}
    \left(
        [\hat P_{i,x},\hat h_j]
        +
        [\hat h_i,\hat P_{j,x}]
    \right).
    \label{eq:cJ_Ph}
\end{align}
The Fourier transform of $\hat P_{i,x}$ is described by the kernel
\begin{align}
    B_x(\vec k_1,\vec k_2)
    =
    \frac{i}{4}
    \left[
        \partial_{k_{1x}}\mathcal H_0(\vec k_1)
        -
        \partial_{k_{2x}}\mathcal H_0(\vec k_2)
    \right].
    \label{eq:Bkernel}
\end{align}
Applying the same procedure as for $\hat J_{y,\mathrm{loc}}^E$, we obtain
\begin{align}
    \mathcal C_{yx}^{(c)}(\vec k,\vec q)
    =
    \frac{1}{2}
    \int_{-1/2}^{1/2}d\lambda\,
    \partial_{p_y}
    \Big[
        B_x(\vec k_-,\vec p)D(\vec p,\vec k_+) 
        \nonumber \\ 
        +
       \left. D(\vec k_-,\vec p)B_x(\vec p,\vec k_+)
    \Big]
    \right|_{\vec p=\vec k+\lambda\vec q}.
    \label{eq:Cc_momentum}
\end{align}
The second contribution of Eq.~\eqref{eq:Cyx_split} can be obtained similarly to the first contribution. Together with the displacement $\delta_{ij,y}$ already present in the current operator, it generates two derivatives with respect to the intermediate momentum. Following the same steps, we find
\begin{align}
    \mathcal C_{yx}^{(\lambda)}(\vec k,\vec q)
   & =
    -i
    \int_{-1/2}^{1/2}d\lambda \lambda\, \nonumber \\ & \times
    \left.
    \partial_{p_x}\partial_{p_y}
    \left[
        D(\vec k_-,\vec p)
        D(\vec p,\vec k_+)
    \right]
    \right|_{\vec p=\vec k+\lambda\vec q}.
    \label{eq:Clambda_momentum}
\end{align}

Combining Eqs.~\eqref{eq:Jloc_momentum}, \eqref{eq:Cc_momentum}, and \eqref{eq:Clambda_momentum}, the flat-space energy-current density operator in Eq.~\eqref{eq:energycurrentoperatorappendix} can be expressed as
\begin{align}
    \hat J_y^E(\vec r)
    =
    \frac{1}{L_xL_y}
    \sum_{\vec k,\vec q}
    e^{i\vec q\cdot\vec r}
    \Psi_{\vec k_-}^\dagger
    \mathcal J_y^E(\vec k,\vec q)
    \Psi_{\vec k_+},
    \label{eq:Jenergy_real_final}
\end{align}
with the energy-current vertex
\begin{align}
    \mathcal J_y^E(\vec k,\vec q)
    =
    \mathcal J_{y,\mathrm{loc}}^E(\vec k,\vec q)
    -
    iq_x
    \left[
        \mathcal C_{yx}^{(c)}(\vec k,\vec q)
        +
        \mathcal C_{yx}^{(\lambda)}(\vec k,\vec q)
    \right].
    \label{eq:Jenergy_vertex}
\end{align}
For the long-wavelength deformation, we expand the energy-current vertex up to third order
\begin{align}
    \mathcal J_y^E(\vec k,q_x)
    \simeq 
    j_0^E(\vec k)
    +
    q_x j_1^E(\vec k)
    +
    q_x^2 j_2^E(\vec k)
    +
    q_x^3 j_3^E(\vec k)\,. 
    \label{eq:Jenergy_qexpansion}
\end{align}
Expanding Eqs.~\eqref{eq:Jloc_momentum}, \eqref{eq:Cc_momentum}, and \eqref{eq:Clambda_momentum} in powers of $q_x$ and using the shorthand notations in Eqs.~\eqref{eq:Hderivatives}, we obtain
\begin{subequations}
\begin{align}
    j_0^E
    &=
    \frac{1}{2}
    \left\{
        \mathcal H,\mathcal H_y
    \right\},
    \label{eq:j0_final}
    \\
    j_1^E
    &=
    \frac{1}{8}
    \left(
        [\mathcal H_y,\mathcal H_x]
        +
        [\mathcal H,\mathcal H_{xy}]
    \right),
    \label{eq:j1_final}
    \\
    j_2^E
    &=
    -\frac{1}{48}
    \left\{
        \mathcal H,\mathcal H_{xxy}
    \right\}
    -
    \frac{5}{96}
    \left\{
        \mathcal H_x,\mathcal H_{xy}
    \right\}
    -
    \frac{1}{96}
    \left\{
        \mathcal H_{xx},\mathcal H_y
    \right\},
    \label{eq:j2_final}
    \\
    j_3^E
    &=
    \frac{1}{192}
    [\mathcal H,\mathcal H_{xxxy}]
    +
    \frac{1}{96}
    [\mathcal H_x,\mathcal H_{xxy}]
    +
    \frac{1}{192}
    [\mathcal H_{xx},\mathcal H_{xy}].
    \label{eq:j3_final}
\end{align}
\end{subequations}
For the two-band Haldane Hamiltonian with crystal orientation $\theta$, 
\begin{align}
    \mathcal H_0(\vec k_\theta) = \mathcal{H}
    =
    \vec d_0(\vec k_\theta)\cdot\vec\sigma,
\end{align}
Eqs.~\eqref{eq:j0_final}--\eqref{eq:j3_final} can be written more explicitly as
\begin{subequations}
\begin{align}
    j_0^E
    &=
    \left(
        \vec d_0\cdot\vec d_{0,y}
    \right)\mathbbm{1},
    \label{eq:j0_d}
    \\
    j_1^E
    &=
    \frac{i}{4}
    \left(
        \vec d_{0,y}\times\vec d_{0,x}
        +
        \vec d_0\times\vec d_{0,xy}
    \right)\cdot\vec\sigma,
    \label{eq:j1_d}
    \\
    j_2^E
    &=
    -\frac{1}{48}
    \left[
        2\vec d_0\cdot\vec d_{0,xxy}
        +
        5\vec d_{0,x}\cdot\vec d_{0,xy}
        +
        \vec d_{0,xx}\cdot\vec d_{0,y}
    \right]\mathbbm{1},
    \label{eq:j2_d}
    \\
    j_3^E
    &=
    \frac{i}{96}
    \big[
        \vec d_0\times\vec d_{0,xxxy}
        +
        2\vec d_{0,x}\times\vec d_{0,xxy}
        \nonumber \\ 
        & \qquad \qquad \qquad \qquad \qquad +
        \vec d_{0,xx}\times\vec d_{0,xy}
    \big]\cdot\vec\sigma.
    \label{eq:j3_d}
\end{align}
\end{subequations}
Here, subscripts on $\vec d_0$ denote momentum derivatives in the same way as for $\mathcal H$, e.g., $\vec d_{0,xy}\equiv \partial_{k_x}\partial_{k_y}\vec d_0(\vec k_\theta)$.

\section{Energy-current response}
\label{app:energycurrent_response}

We first consider the equilibrium and contact contributions to the energy-current response. As in the charge-current response, the frequency integration projects the equilibrium Green function onto the occupied band. Since the equilibrium Green function is diagonal in momentum, only the zero-momentum-transfer component of the energy-current vertex, $q_x=0$, contributes. At $q_x=0$, the energy-current vertex $j_0^E$ in Eq.~\eqref{eq:j0_d} is simply
\begin{align}
    j_0^E
    =
    \frac{1}{2}
    \partial_{k_y}
    |\vec d_0|^2\,\mathbbm 1.
    \label{appeq:energycurrent_q0}
\end{align}
The equilibrium energy current therefore vanishes,
\begin{align}
    J_{y,\mathrm{eq}}^E
    &=
    \int_{\rm BZ}\frac{d^2k}{(2\pi)^2}
    \Tr\left[
        P_-(\vec k_\theta)j_0^E
    \right]
    =0,
    \label{appeq:energycurrent_equilibrium}
\end{align}
where the last equality follows from the periodicity of the Brillouin zone and holds for arbitrary crystal orientation $\theta$.

We next consider the contact contribution from the deformation-induced correction to the energy-current vertex. Since it is contracted with the equilibrium Green function, only the $q_x=0$ correction contributes. For the scalar-potential deformation,
\begin{align}
    \delta\mathcal J_{y,\Phi}^E(\vec k,0,x;\theta)
    =
    \Phi(x)\,
    \vec d_{0,y}\cdot\vec\sigma,
    \label{appeq:energycurrent_contact_phi}
\end{align}
while the hopping deformation gives
\begin{align}
    \delta\mathcal J_{y,\tau}^E(\vec k,0,x;\theta)
    =
    \partial_{k_y}
    \left[
        \vec d_0\cdot
        \delta\vec d(\vec k,x;\theta)
    \right]\mathbbm 1.
    \label{appeq:energycurrent_contact_tau}
\end{align}
The contact contribution therefore becomes
\begin{align}
    J_{y,\mathrm{cont}}^E(x;\theta)
    &=
    \int_{\rm BZ}\frac{d^2k}{(2\pi)^2}
    \Tr\left[
        P_-(\vec k_\theta)
        \delta\mathcal J_y^E(\vec k,0,x;\theta)
    \right]
    \nonumber\\
    &=
    \int_{\rm BZ}\frac{d^2k}{(2\pi)^2}
    \Big\{
        \Phi(x)
        \partial_{k_y}E_-(\vec k_\theta)
        \nonumber\\
    &\qquad\qquad
        +
        \partial_{k_y}
        \left[
            \vec d_0\cdot
            \delta\vec d(\vec k,x;\theta)
        \right]
    \Big\}
    =0.
    \label{appeq:energycurrent_contact_zero}
\end{align}
Thus, both the equilibrium and contact contributions vanish as total derivatives over the Brillouin zone for arbitrary crystal orientation $\theta$.

We now consider the bubble contribution. Since the energy-current vertex itself depends on the transferred momentum $q_x$, both the current vertex and the Green functions must be expanded. In the following, the orientation-angle dependence is left implicit. For a deformation vertex $\Gamma=\Phi,\tau_\ell$, corresponding to the scalar-potential and hopping deformations, respectively, we define the energy-current bubble kernel by
\begin{align}
    \mathcal K_{\Gamma}^E(\vec k_\theta,q_x)
    =
    -2\int\frac{d\omega}{2\pi}f_0(\omega)
    \operatorname{Im}
    \Tr\left[
        \mathcal J_y^E(\vec k_\theta,q_x)
        g_{\vec k_+}^R
        \Gamma
        g_{\vec k_-}^R
    \right].
    \label{appeq:energy_bubble_kernel}
\end{align}
Using the expansions, 
\begin{align}
    \mathcal J_y^E(\vec k_\theta,q_x)
    &\simeq
    j_0^E
    +
    q_x j_1^E
    +
    q_x^2 j_2^E
    +
    q_x^3 j_3^E,
    \\
    g_{\vec k_+}^R\Gamma g_{\vec k_-}^R
    &\simeq
    B_0^\Gamma
    +
    q_x B_1^\Gamma
    +
    q_x^2 B_2^\Gamma
    +
    q_x^3 B_3^\Gamma,
\end{align}
we expand $\mathcal K_{\Gamma}^E(\vec k_\theta,q_x)$ up to third order 
\begin{align}
    \mathcal K_{\Gamma}^E(\vec k_\theta,q_x)
    \simeq 
    \sum_{n=0}^{3}
    q_x^n\mathcal K_{\Gamma}^{E,(n)}(\vec k_\theta).
    \label{appeq:energy_kernel_expansion}
\end{align}
The coefficients up to cubic order are
\begin{subequations}
\begin{align}
    \mathcal K_{\Gamma}^{E,(0)}
    &=
    -2\int\frac{d\omega}{2\pi}f_0(\omega)
    \operatorname{Im}\Tr
    \left[
        j_0^E B_0^\Gamma
    \right],
    \\
    \mathcal K_{\Gamma}^{E,(1)}
    &=
    -2\int\frac{d\omega}{2\pi}f_0(\omega)
    \operatorname{Im}\Tr
    \big[
        j_0^E B_1^\Gamma
        +
        j_1^E B_0^\Gamma
    \big],
    \\
    \mathcal K_{\Gamma}^{E,(2)}
    &=
    -2\int\frac{d\omega}{2\pi}f_0(\omega)
    \operatorname{Im}\Tr
    \left[
        j_0^E B_2^\Gamma
        +
        j_1^E B_1^\Gamma
        +
        j_2^E B_0^\Gamma
    \right],
    \\
    \mathcal K_{\Gamma}^{E,(3)}
    &=
    -2\int\frac{d\omega}{2\pi}f_0(\omega)
    \operatorname{Im}\Tr
    \big[
        j_0^E B_3^\Gamma
        +
        j_1^E B_2^\Gamma \nonumber \\ 
        & \qquad \qquad \qquad \qquad \quad 
        +
        j_2^E B_1^\Gamma
        +
        j_3^E B_0^\Gamma
    \big].
\end{align}
\label{appeq:energy_kernel_coefficients}
\end{subequations}
For example, in terms of the frequency-integrated functions introduced in Appendix~\ref{appsubsec:frequencyintegration}, the linear-order kernel is
\begin{align}
    \mathcal K_{\Gamma}^{E,(1)}
    =
    \omega_2(j_1^E,\Gamma)
    +
    \frac{1}{2}
    \left[
        \omega_3(j_0^E,\mathcal H_x,\Gamma)
        -
        \omega_3(j_0^E,\Gamma,\mathcal H_x)
    \right].
    \label{appeq:energy_kernel_linear}
\end{align}
The energy-current response coefficients are obtained by integrating the momentum-resolved kernels over the Brillouin zone, 
\begin{align}
    \kappa_{\Gamma}^{E,(n)}(\theta)
    \equiv
    (-i)^n
    \int_{\rm BZ}\frac{d^2k}{(2\pi)^2}
    \mathcal K_{\Gamma}^{E,(n)}(\vec k_\theta),
    \label{eq:energy_response_coefficient}
\end{align}
Performing the remaining real-space integral and the sum over $q_x$ using
\begin{align}
    \frac{1}{L_x}\sum_{q_x}\int dx_1\,
    e^{iq_x(x-x_1)}q_x^n\Gamma(x_1)
    =
    (-i)^n\partial_x^n\Gamma(x),
    \label{appeq:q_to_gradient_energy}
\end{align}
we obtain the bubble response as 
    \begin{align}
    \delta J_{y,\Gamma}^E(x;\theta)
    =
    \sum_{n=0}^{3}
    \kappa_{\Gamma}^{E,(n)}(\theta)
    \partial_x^n\Gamma(x),
    \label{appeq:energy_response_gradient}
\end{align}
which reproduces Eq.~\eqref{eq:energy_response_gradient} of the main text. 

\bibliography{ribbon}

\end{document}